\documentclass[final,5p,times,twocolumn,authoryear]{elsarticle}

\usepackage{amsmath,amssymb}
\allowdisplaybreaks
\usepackage{xcolor}
\usepackage{algorithm}
\usepackage{algpseudocode}
\usepackage{graphicx}
\usepackage{textcomp}
\usepackage{import}
\usepackage{multirow}
\usepackage{natbib}
\usepackage{tablefootnote}
\usepackage[hidelinks]{hyperref} 

\DeclareMathOperator*{\argmin}{argmin} 

\newtheorem{ass}{Assumption}
\newtheorem{thm}{Theorem}

\newtheorem{rem}{Remark}

\journal{ }

\begin{document}

\begin{frontmatter}

\title{ Cooperative Distributed MPC via Decentralized Real-Time Optimization:\\
Implementation Results for Robot Formations
}
	
\author[TUDO]{Gösta Stomberg}
\ead{goesta.stomberg@tu-dortmund.de}
\affiliation[TUDO]{organization={Institute for Energy Systems, Energy Efficiency and Energy Economics, TU Dortmund University},addressline={44227 Dortmund},country={Germany}}

\author[UStuttgart]{Henrik Ebel}
\ead{henrik.ebel@itm.uni-stuttgart.de}
\author[TUDO]{\corref{cor1}Timm Faulwasser}
\cortext[cor1]{Corresponding author}
\ead{timm.faulwasser@ieee.org}
\author[UStuttgart]{Peter Eberhard}
\ead{peter.eberhard@itm.uni-stuttgart.de}
        
\affiliation[UStuttgart]{organization={Institute of Engineering and Computational Mechanics, University of Stuttgart},     	addressline={70569 Stuttgart},country={Germany}}
      
\begin{abstract}
Distributed model predictive control (DMPC) is a flexible and scalable feedback control method applicable to a wide range of systems.
While the stability analysis of DMPC is quite well understood, there exist only limited implementation results for realistic applications involving distributed computation and  networked communication. 
This article  approaches formation control of mobile robots via a cooperative DMPC scheme. We discuss the implementation via decentralized optimization algorithms.
To this end, we combine the alternating direction method of multipliers with decentralized sequential quadratic programming to solve the underlying optimal control problem in a decentralized fashion with nominal convergence guarantees.
Our approach only requires coupled subsystems to communicate and does not rely on a central coordinator.
Our experimental results showcase the efficacy of DMPC for formation control and they demonstrate the real-time feasibility of the considered algorithms.

\end{abstract}


\end{frontmatter}

\section{Introduction}\label{sec:intro}

Model predictive control~(MPC) is an advanced feedback design, which provides a structured framework for regulation, stabilization, and other control tasks.
Motivated by the success of predictive control in applications, numerous distributed MPC (DMPC) schemes have been proposed for cyber-physical systems with applications ranging from energy systems~\citep{Venkat2008} to robot swarms~\citep{vanParys2017}.
Cooperative distributed control utilizes local sensing, local computation, and networked capabilities of subsystems to replace a centralized controller.
Put differently, the subsystems cooperatively solve the control task, which improves scalability compared to a centralized approach~\citep{Muller2017}.

Robotic swarms are a prime application to study and showcase the intricacies of distributed control and to demonstrate the advantages of cooperative DMPC. 
On the one hand, this is motivated from a theoretical perspective since robotic swarms exhibit many challenges of distributed control---plant-model mismatch, issues of communication, high sampling frequency, as well as highly cooperative and scalable tasks with numerous possible constraints. 
Moreover, physical robotic cooperation is often rather intuitive, making it straight-forward to measure, to analyze, to depict, and to grasp the performance of different distributed schemes. 
On the other hand, progress in this area is highly relevant to engineering applications. 
Well-controlled swarms of land-based and aerial mobile robots are expected to revolutionize logistics, transportation, and security~\citep{GrauEtAl17,WenHeZhu18}. 

\begin{figure*}
	\centering
	\includegraphics[width=\textwidth]{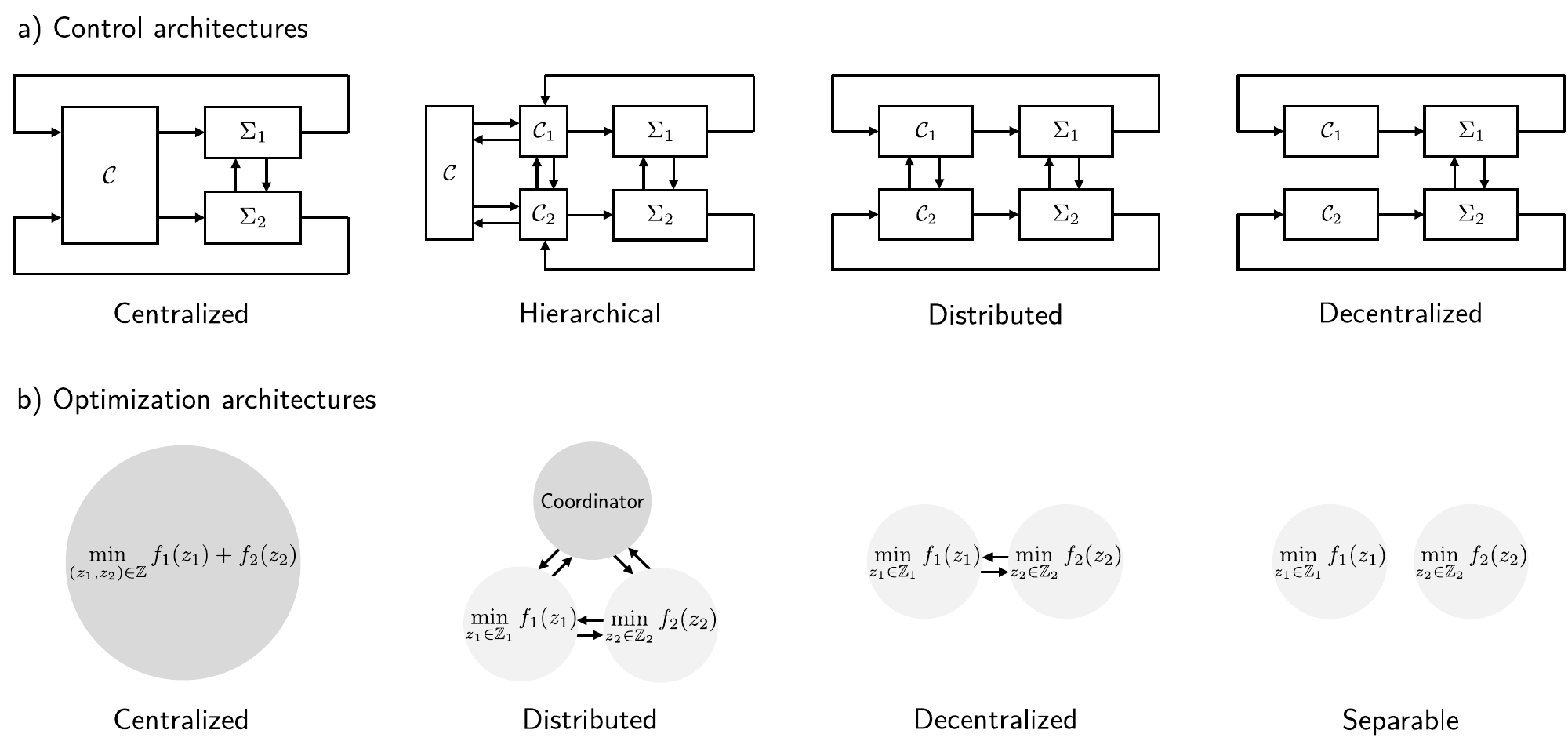}
	\caption{Control architectures and corresponding optimization architectures for MPC. This article discusses DMPC via decentralized optimization.}\label{fig:schemes}
\end{figure*}

The present article considers formation control of mobile robots, which is one of the fundamental tasks in distributed robotics~\citep{Olfati-Saber2007,BaiArcakWen11}. Specifically,  formation control can serve as a building block for many relevant applications, such as cooperative transportation and manipulation~\citep{EbelEberhard22}. 
Moreover, formation control has served as an example for the development and evaluation of DMPC. 
For instance, \cite{YanBitmead03} present one of the early applications of MPC for the formation control of vehicles, although they focus on one-dimensional formations do not validate the scheme on hardware. 
\cite{GuYang05} consider a centralized optimal control problem~(OCP) for two-dimensional formations, but conducted no hardware experiments. 
In contrast, \cite{KanjanawanishkulZell08} tested an MPC-based formation controller on real-world omnidirectional mobile robots where each robot runs its own controller. 
However, there is a centralized leader agent and the remaining robots merely follow the leader by using a conventional MPC controller to track the trajectory predicted by the leader, which does not fit today's notion of cooperative DMPC where robots jointly solve an optimization problem. 
Yet, \cite{KanjanawanishkulZell08} as well as \cite{KuwataEtAl07} seem to be among the first to deploy DMPC with distributed computation and communication on hardware. 
Since then, formation control has remained a popular benchmark problem accompanying the theoretical and numerical progress of DMPC. 
For instance, \cite{EbelSharafianArdakaniEberhard17} compare sequential and cooperative DMPC methods for linear systems and analyze the performance of both schemes for a formation control task with hardware. 
Similarly, \cite{Ebel2021} use formation control to compare a cooperative DMPC scheme to a distributed control method based on algebraic graph theory, confirming advantages of DMPC with respect to constraint handling also in hardware experiments. 
Recently, \cite{Burk2021b} tested the performance of an open-source DMPC framework with formation control, including real-world hardware but without distributed computation in the experiments. 
\cite{NovothEtAl21} study more complex scenarios including obstacle avoidance, but without the consideration of hardware experiments and distributed computation. 
Often in formation control, omnidirectional mobile robots are considered, whereas \cite{RosenfelderEbelEberhard22} tailor a DMPC controller also to mobile robots with non-holonomic constraints.

From the systems and control point of view, one can distinguish different architectures ranging from centralized via distributed to decentralized schemes, cf.  Figure~\ref{fig:schemes}\,a).
However, the notions for distributed and decentralized architectures differ slightly in control and optimization, cf. Figure~\ref{fig:schemes}\,b).
Specifically, in centralized feedback architectures, a single  unit~$\mathcal{C}$ controls all subsystems $\Sigma_i$.
If~$\mathcal{C}$ is an MPC controller, then a centralized OCP is solved via so-called centralized optimization algorithms.
If the OCP is solved such that the computation is distributed onto subsystems that communicate both with one another and with a coordinator, then one speaks of a hierarchically distributed architecture. 
Finally, in distributed control, there exists a local controller $\mathcal{C}_i$ for each subsystem $\Sigma_i$ and these controllers communicate only with other controllers, but not with a central coordinator.
To obtain such a distributed architecture, the OCP needs to be solved via decentralized algorithms which only require subsystem-to-subsystem communication.
Finally, in decentralized control, each subsystem~$\Sigma_i$ is controlled by a controller~$\mathcal{C}_i$ and there is no communication between the controllers.
This corresponds to separable OCPs which are solved by each subsystem individually without any communication. 
We refer to~\citep{Scattolini2009} for an overview of MPC architectures and to~\citep{Bertsekas1989,Nedic2018a} for the taxonomy of distributed and decentralized optimization. 
In order to tackle the robot formation control problem,  this article focuses on cooperative DMPC, where the underlying centralized OCP is solved iteratively via decentralized optimization algorithms.
That is, we implement a \emph{distributed} controller through \emph{decentralized} optimization.

In principle, the chosen approach of solving a centralized OCP via decentralized optimization can achieve a performance equivalent to a centralized MPC controller.
In practice, the performance of DMPC is likely slightly worse than the one of centralized MPC, because the number of optimizer iterations per sampling interval is limited.
There exists a range of decentralized optimization methods applicable to DMPC and we refer to~\citep{Stomberg2022} for a recent overview for linear-quadratic problems.
In particular the alternating direction method of multipliers~(ADMM) shows promising performance in numerical studies~\citep{Conte2012a,Rostami2017}. 
ADMM is guaranteed to converge for convex problems~\citep{Boyd2011}, but lacks convergence guarantees for general non-convex constraints arising from nonlinear dynamics or otherwise.
A decentralized method with convergence guarantees also for non-convex problems is the recently proposed decentralized sequential quadratic programming~(dSQP) method, which combines an SQP scheme with ADMM~\citep{Stomberg2022a} in a bi-level fashion.

The design, numerical implementation, and stability of cooperative DMPC are addressed in~\citep{Stewart2011,Giselsson2014,Conte2016,Kohler2019,Bestler2019}.
\cite{Stewart2011} propose a tailored optimization method where each subsystem is equipped with the models of all subsystems such that the optimization method generates feasible iterates. 
Stability then follows from the design of terminal ingredients in the OCP. 
\cite{Giselsson2014} and~\cite{Kohler2019} guarantee stability through constraint tightening when applying dual decomposition to OCP designs with and without terminal ingredients, respectively.
\cite{Bestler2019} develop a stopping criterion for ADMM to guarantee stability under the assumption that ADMM converges linearly for non-convex OCPs.

However, there exist only few experimental results to validate the developed DMPC schemes under real-world conditions.
Therefore, we focus on the implementation, numerics, and experimental testing with real robot hardware.
\cite{vanParys2017} and~\cite{Burk2021b} made notable contributions to the experimental validation of cooperative DMPC.
The former tested an ADMM-based DMPC scheme on non-holonomic mobile robots with nonlinear dynamics.
The DMPC algorithm was implemented on embedded computers onboard the robots such that there was one computing entity per robot. 
While the control performance was satisfactory, only a single ADMM iteration could be applied per control step due to the high computational complexity of the nonlinear program (NLP) each robot had to solve online.
Even with infinitely many iterations, the application of ADMM to problems with non-convex constraint sets does not admit general convergence guarantees.
\cite{Burk2021b} tested the open-source framework {GRAMPC-D} for the cooperative control of nonlinear unicycles running multiple ADMM iterations per control step. 
However, the test setup did not allow for true distributed execution as all control algorithms were implemented on a single computer.

This article focuses on the practical application of available DMPC methods and presents experimental results for ADMM- and dSQP-based DMPC schemes for a team of omnidirectional mobile robots. 
We use ADMM~\citep{Boyd2011} if the OCP transfers into a convex quadratic program (QP), and we use dSQP~\citep{Stomberg2022a} if a non-convex NLP is obtained.
In both cases, only QPs are solved on a subsystem level, which allows to use the efficient active set method \texttt{qpOASES} tailored to MPC~\citep{Ferreau2014}.
Our decentralized and efficient implementation of ADMM- and dSQP-based DMPC allows to run multiple optimizer iterations per control step.
We assign a single computer to each robot and thus we can test the algorithms in a distributed control setting.

The main contribution of this article is the practical evaluation of a decentralized optimization algorithm for DMPC. Specifically and to the best of the authors' knowledge, the present paper appears to be the first to present experimental implementation results for cooperative distributed MPC with non-convexity in the underlying OCPs \textit{and} with nominal convergence guarantees of the employed algorithm.

The paper is structured as follows. 
Section~\ref{sec:problem} introduces the DMPC scheme and the OCP. 
Section~\ref{sec:opt} presents ADMM and dSQP for solving the optimization problem. 
Section~\ref{sec:implementation} gives an overview of implementational aspects with regard to the algorithms and experimental realization. 
Section~\ref{sec:exp} analyses the performance and execution time of the DMPC schemes and gives valuable insights for the future development of distributed optimization algorithms tailored to the needs of DMPC. 

\textit{Notation:} 
Given a matrix $A$ and an integer $j$, $[A]_j$ denotes the $j$th row of $A$.
For an index set $\mathcal{A}$, $[A]_\mathcal{A}$ denotes the matrix consisting of rows $ [A]_j$ for all $j \in \mathcal{A}$.
Likewise, $[a]_j$ is the $j$th component of vector $a$ and $a_\mathcal{A}$ is the vector of components $[a]_j$ for all $j \in \mathcal{A}$.
The concatenation of vectors $x$ and $y$ into a column vector is $\text{col}(x,y)$.
Likewise, $\text{col}(x_i)_{i \in \mathcal{S}}$ denotes the concatenation of vectors $x_i$ for all $i \in \mathcal{S}$.
The $\varepsilon$-neighborhood around a point $x \in \mathbb{R}^n$ is denoted by $\mathcal{B}_\varepsilon(x)$, i.e., ${\mathcal{B}_\varepsilon(x) \doteq \{y \in \mathbb{R}^{n}\ \vert \ \| y - x\| \leq \varepsilon\}}$, where $\| \cdot \|$ is some norm on $\mathbb{R}^n$.
Further, $A = [A_{ij}]$ is the block matrix with entries $A_{ij}$ at block position $(i,j)$. 
$I$ is the identity matrix.
The set $\mathbb{I}_{[0,N]}$ denotes the integers in the range $[0,N]$.
The time is denoted by $t$.

\section{Problem Statement}\label{sec:problem}

Let $\mathcal{S} = \{1,\dots, S\}$ be a set of mobile robots. 
The state and input of robot $i \in \mathcal{S}$ are denoted as $x_i \in \mathbb{R}^{n_{x,i}}$ and $u_i \in \mathbb{R}^{n_{u,i}}$.
In this article, the state $x_i$ represents the position of robot $i$ with respect to a fixed coordinate system, and the input $u_i$ represents the desired translational velocity of the robot body. 
The centralized position and velocity of all robots are $x \doteq \text{col}(x_i)_{i \in \mathcal{S}} \in \mathbb{R}^{n_x}$ and $u \doteq \text{col}(u_i)_{i \in \mathcal{S}} \in \mathbb{R}^{n_u}$.

\subsection{DMPC of Mobile Robots}

\begin{figure}
	\centering
	\includegraphics[width=0.35\textwidth]{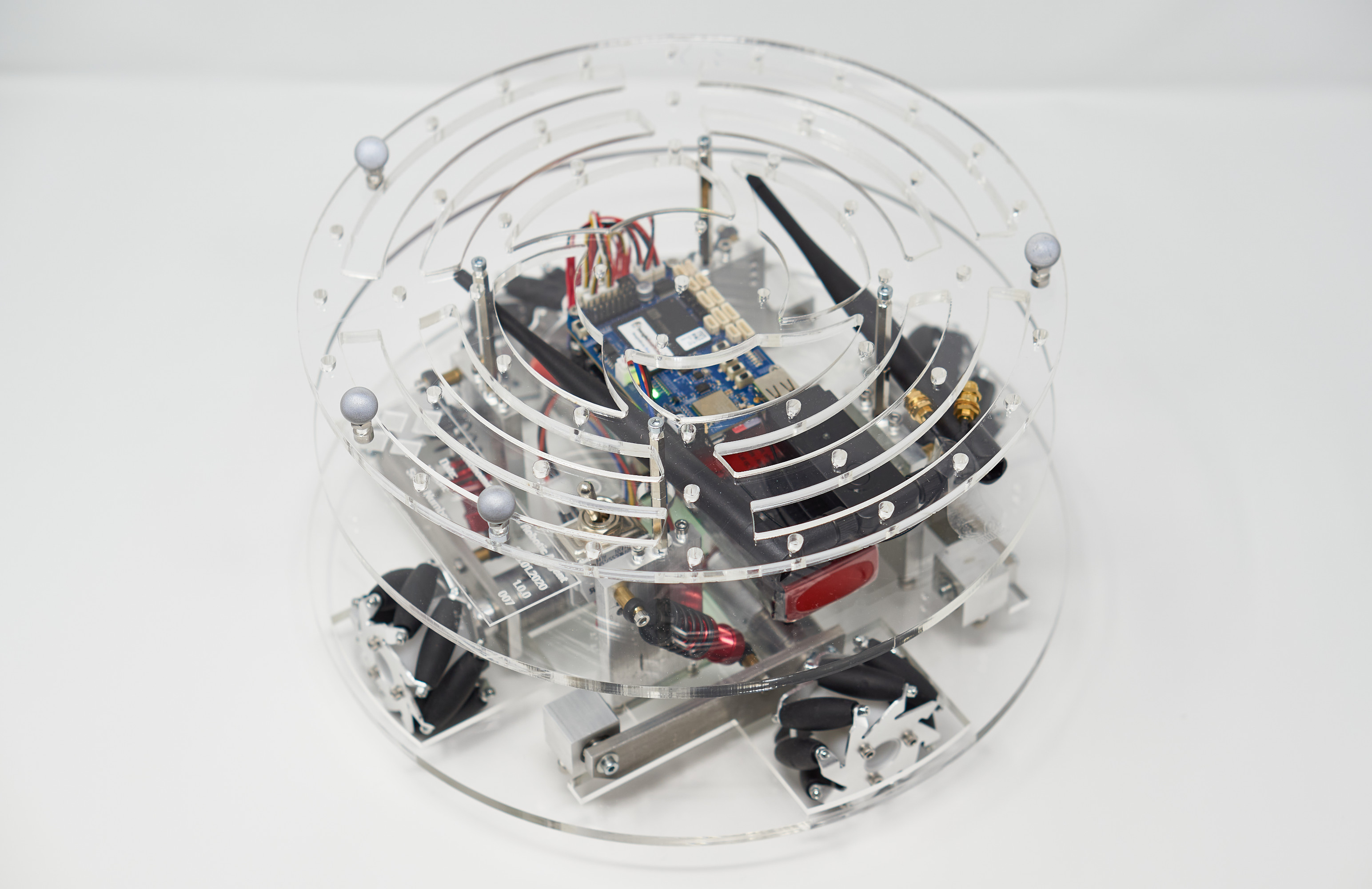}
	\caption{Omnidirectional mobile robot for distributed robotics experiments.}
	\label{fig:hera}
\end{figure}

We consider the task where the centralized position $x$ tracks a desired formation $\bar{x}(t) \doteq \text{col}(\bar{x}_i(t))_{i \in \mathcal{S}}$.
Figure~\ref{fig:hera} shows one of the custom mobile robots we use for the experimental validation of our approach~\citep{EbelEberhard22}. 
The omnidirectional robots move in the $X$-$Y$-plane and they can accelerate into any direction at any time, independent from their orientation. 
The (approximate) dynamics of each robot used for prediction in the controller are given by ${f_i^\mathrm{d}(x_i,u_i) = x_i + \Delta t\, u_i}$, where $\Delta t$ denotes the control sampling interval and where ${n_{x,i} = n_{u,i} = 2}$.

\subsection{OCP Formulation}
To design the DMPC scheme, we first formulate a centralized~OCP
\begin{subequations}\label{ocp}
	\begin{align}
	\min_{\boldsymbol{x},\boldsymbol{u}} &\sum_{i \in \mathcal{S}} \sum_{k=0}^{N-1} \ell_i (x^k,u_i^k) + V_{\text{f},i}(x^N)\label{ocp:cost}\\
	\nonumber\text{subject} &\text{ to for all } i \in \mathcal{S}\\
	x_i^{k+1} &= f_i^\mathrm{d}(x_i^k,u_i^k), && \forall k \in \mathbb{I}_{[0,N-1]},\label{ocp:dyn}\\
	(x_i^0,u_i^0) &= (x_i(t),u_i(t)),\label{ocp:x0}\\
	x_i^k &\in \mathbb{X}_i && \forall k \in \mathbb{I}_{[0,N]}\label{ocp:xi},\\
	u_i^k &\in \mathbb{U}_i && \forall k \in \mathbb{I}_{[0,N-1]}\label{ocp:ui},\\
	(x_i^k,x_j^k) &\in \mathbb{X}_{ij}, \hspace{0.49cm} \forall j \in \mathcal{S}, && \forall k \in \mathbb{I}_{[0,N]}.\label{ocp:xij}
	\end{align}
\end{subequations}

The decision variables $\boldsymbol{x} \doteq \text{col}(x^k)_{k \in \mathbb{I}_{[0,N]}}$ and $\boldsymbol{u} \doteq \text{col}(u^k)_{k \in \mathbb{I}_{[0,N-1]}}$ are the predicted trajectories over the horizon~$N$.
OCP~\eqref{ocp} couples the robots via the stage cost ${\ell_i : \mathbb{R}^{n_{x,1}} \times \dots \times \mathbb{R}^{n_{x,S}} \times \mathbb{R}^{n_{u,i}} \rightarrow \mathbb{R}}$,
\begin{align*}
\ell_i(x,u_i) \doteq  \sum_{j \in \mathcal{S}} \frac{1}{2} (x_i-\bar{x}_i)^\top Q_{ij} (x_j-\bar{x}_j) + \frac{1}{2} (u_i - \bar{u}_i)^\top R_{ii} (u_i - \bar{u}_i), 
\end{align*} via the terminal penalty $V_{\text{f},i} : \mathbb{R}^{n_{x,1}} \times \dots \times \mathbb{R}^{n_{x,S}} \rightarrow \mathbb{R}$,
\begin{align*}
V_{\text{f},i}(x) \doteq \frac{1}{2} \sum_{j \in \mathcal{S}} (x_i-\bar{x}_i)^\top P_{ij} (x_j-\bar{x}_j),
\end{align*} where $Q_{ij} \in \mathbb{R}^{n_{x,i} \times n_{x,j}}$, $R_{ii} \in \mathbb{R}^{n_{u,i} \times n_{u,i}} $, and $P_{ij} \in \mathbb{R}^{n_{x,i} \times n_{x,j}}$ are weight matrices, and via the constraints~\eqref{ocp:xij}, where $\mathbb{X}_{ij} \subseteq  \mathbb{R}^{n_{x,i}}  \times \mathbb{R}^{n_{x,j}}$.
The sets $\mathbb{X}_i \subseteq \mathbb{R}^{n_{x,i}} $ and $\mathbb{U}_i \subseteq \mathbb{R}^{n_{u,i}} $ constrain the states and inputs of subsystem $i$, respectively.
The coupling in OCP~\eqref{ocp} describes the cooperation between the robots to achieve the control task.

We only consider state coupling in OCP~\eqref{ocp} and refer to~\citep{Stomberg2022} for an extension to input coupling.
The coupling structure of OCP~\eqref{ocp} can be represented as a graph and we define for each robot~$i$ sets of in- and out-neighbors
\begin{align*}
\mathcal{N}_i^\text{in} &\doteq \left\{j \in \mathcal{S}\backslash \{i\}   \ \vert\ Q_{ij} \neq 0 \cup P_{ij} \neq 0 \cup \mathbb{X}_{ij} \neq \mathbb{R}^{n_{x,i}} \times \mathbb{R}^{n_{x,j}} \right\},\\
\mathcal{N}_i^\text{out} &\doteq \left\{ j \in \mathcal{S} \ \vert\ i \in \mathcal{N}_j^\text{in}\right\}.
\end{align*} 
We make the following two assumptions on the communication between subsystems and on OCP~\eqref{ocp}.
\begin{ass}[Communication graph]
	We assume that the communication graph is equivalent to the coupling graph, i.e., each subsystem $i$ communicates bi-directionally with all its neighbors $\mathcal{N}_i \doteq \mathcal{N}_i^\text{in} \cup \mathcal{N}_i^\text{out}$ to solve OCP~(1) cooperatively.
	We further assume that no messages are lost between subsystems.
\end{ass}
\begin{ass}[OCP~\eqref{ocp} requirements]\phantom{m}\label{ass:ocp}
	\begin{enumerate}
		\item The centralized weight matrices $Q \doteq [Q_{ij}] \in \mathbb{R}^{n_x \times n_x}$,\\ $R \doteq [R_{ij}] \in \mathbb{R}^{n_u \times n_u}$, and $P \doteq [P_{ij}] \in \mathbb{R}^{n_x \times n_x}$ are symmetric positive definite.
		\item The sets $\mathbb{X}_i$, $\mathbb{U}_i$, and $\mathbb{X}_{ij}$ are closed for all $i,j\in \mathcal{S}$.\hfill $\square$
	\end{enumerate}	
\end{ass}

Most approaches in the literature guarantee closed-loop stability based on Assumption~\ref{ass:ocp} combined with additional assumptions on the compactness of the constraint sets $\mathbb{X}_i$ and $\mathbb{U}_i$, the design of terminal constraints and terminal penalties, and initial feasibility of the OCP.
Here, we omit the design of specific terminal ingredients and instead focus on the numerical implementation and the experimental validation. Instead we refer to \citep{Conte2016} and \citep{Darivianakis2019} for the design of terminal ingredients in DMPC.
Moreover, we allow for a non-convex set $\mathbb{X}_i$ as this enables guaranteeing collision avoidance via encoding minimum-distance requirements between robots in tailored constraints.

Algorithm~\ref{alg:DMPC} summarizes the considered DMPC scheme. 
Per MPC step, each robot first measures its current position $x_i(t)$.
Then, the robots solve OCP~\eqref{ocp} cooperatively using a decentralized optimization method.
The second element $u_i^1$ of the optimal open-loop input trajectory $\boldsymbol{u}_i^\star$ is then chosen as input $u_i(t+\Delta t)$ in the next MPC step.

\begin{rem}[Compensation of computational delay]\phantom{The}	
The careful reader will notice that~\eqref{ocp:x0} imposes an initial condition on the input.
This choice allows to compensate the computational delay of Steps 3 and 4 of Algorithm~\ref{alg:DMPC}.
Solving the OCP in decentralized fashion can take a substantial fraction of the samping interval due to the iterative nature of decentralized optimization methods and due to communication latencies between robots. 
Applying $u_i^1$ in the next MPC step instead of applying $u_i^0$ in the current MPC step compensates for the inevitable computational delay~\citep{Findeisen2006}. This approach therefore requires the combined execution time for Steps 3--5 of Algorithm~\ref{alg:DMPC} to be less than $\Delta t$. \hfill $\square$
\end{rem}

\begin{algorithm}[t]
	\caption{Cooperative distributed MPC} \label{alg:DMPC}
	\begin{algorithmic}[1]
		\State Initialization: $u_i(0), \bar{x}_i$ for all $i \in \mathcal{S}$
		\While{$t < t_{\text{end}} $ for all $i \in \mathcal{S}$}
		\State Measure $x_i(t)$ for all $i \in \mathcal{S}$	
		\State Solve OCP~\eqref{ocp} via decentralized optimization
		\State Extract $u_i^1$ from $\boldsymbol{u}_i^\star$  
		\State $u_i(t+\Delta t) = u_i^1$
		\State $t = t+\Delta t$
		\EndWhile\label{euclidendwhile}
	\end{algorithmic}
\end{algorithm}

\subsection{Reformulation as Partially Separable Problem}
To obtain a DMPC scheme, we solve OCP~\eqref{ocp} online via iterative decentralized optimization methods. 
To this end, we introduce state copies $w_{ji}^k = x_j^k$ for all $j \in \mathcal{N}_i^\text{in}$ and for all $i \in \mathcal{S}$.
The trajectory copies $\boldsymbol{w}_i \doteq \text{col}(w_i^k)_{k \in \mathbb{I}_{[0,N]}}$ with $w_i^k \doteq \text{col}(w_{ji}^k)_{j \in \mathcal{N}_i^{\text{in}}}$ are additional decision variables of subsystem $i$ and $w_i^k$ replaces $x_j^k$ in \eqref{ocp:cost}, \eqref{ocp:dyn}, and \eqref{ocp:xij} for all $j \in \mathcal{N}_i^{\text{in}}$.
We reformulate OCP~\eqref{ocp} with $z_i \doteq \text{col}(\boldsymbol{x}_i,\boldsymbol{u}_i,\boldsymbol{w}_i)\in \mathbb{R}^{n_i}$ as the partially separable NLP
\begin{subequations}\label{nlp}
	\begin{align}
	\min_{z_i \in \mathbb{R}^{n_i},i \in \mathcal{S}} &\sum_{i \in \mathcal{S}} f_i (z_i)\label{nlp:obj}\\
	\nonumber\text{subject} &\text{ to}\\
	g_i(z_i) &=0 \quad \;\vert\; \nu_i, \quad \forall i \in \mathcal{S},\label{nlp:eq}\\ 
	h_i(z_i) &\leq 0 \quad \;\vert\; \mu_i, \quad \forall i \in \mathcal{S},\label{nlp:ineq}\\ 
	\sum_{i \in \mathcal{S}} E_i z_i &= 0 \quad \;\vert\; \lambda, \label{nlp:coup}
	\end{align}
\end{subequations}
where $f_i : \mathbb{R}^{n_i} \rightarrow \mathbb{R}$, $g_i : \mathbb{R}^{n_i} \rightarrow \mathbb{R}^{n_{g,i}}$, and $h_i : \mathbb{R}^{n_i} \rightarrow \mathbb{R}^{n_{h,i}}$ are twice continuously differentiable. 
The notation in~\eqref{nlp} highlights that $\nu_i$, $\mu_i$, and $\lambda$ are the Lagrange multipliers to \eqref{nlp:eq}, \eqref{nlp:ineq}, and \eqref{nlp:coup}. 
The objective ${f_i(z_i) \doteq 1/2\, z_i^\top H_i z_i + g_i^\top z_i}$ of subsystem $i$ represents~\eqref{ocp:cost} and the matrices $H_i$ are composed of the weight matrices $Q_{ij},R_{ii}$, and $P_{ij}$.
The equality constraint~\eqref{nlp:eq} includes the system dynamics~\eqref{ocp:dyn} and the initial condition~\eqref{ocp:x0}. Depending on the type of constraint,~\eqref{ocp:xi}--\eqref{ocp:xij} are part of~\eqref{nlp:eq} or~\eqref{nlp:ineq}. 
The constraint~\eqref{nlp:coup} couples states and copies, see \citep[Example 1]{Stomberg2022}.

\begin{rem}[Soft-constrained optimal control problem]\label{rem:softOCP}
\phantom{We}
Since we omit the design of tailored terminal penalties and terminal constraints, OCP~\eqref{ocp} is not guaranteed to be recursively feasible in the presence of state constraints.
Therefore, we introduce a scalar slack variable $s_i \geq 0$ as an additional decision variable for each robot, cf. \citep[Eq. 3.91]{Maciejowski2002}.
We then replace any hard inequality constraint $[h_i(z_i)]_p$ in~\eqref{nlp:ineq} corresponding to a state constraint by the soft inequality constraint $[h_i(z_i)]_p \leq s_i$.
The slack $s_i$ is penalized by adding $c s_i^2$ with $c \gg 0$ to~\eqref{nlp:obj}.
Note that the soft-constrained problem obtained this way is always feasible. \hfill $\square$

\end{rem}

\section{Decentralized Optimization}\label{sec:opt}

The key concept to implement Algorithm~\ref{alg:DMPC} as a \textit{distributed} controller lies in the solution of the centralized OCP using \textit{decentralized} optimization methods. 
That is, we use methods that only require robots that are neighbors in the underlying graph to communicate with each other for solving the OCP.
Specifically, this section discusses an ADMM variant and dSQP. 

\subsection{Essentials of ADMM}

To apply ADMM, we introduce the auxiliary decision variable $\bar{z}_i \in \mathbb{R}^{n_{i}}$ for each subsystem and rewrite~\eqref{nlp} as
\begin{subequations}
	\label{ocp:admm}
	\begin{align}
	\min_{z_i \in \mathbb{Z}_i, \bar{z}_i \in \mathbb{R}^{n_{i}}, i \in \mathcal{S}} \quad  &\sum_{i\in \mathcal{S}} f_i(z_i)
	\\
	\nonumber \textrm{subject} &\textrm{ to}\\
	z_i - \bar{z}_i &= 0 \quad \; | \;  \gamma_i, \quad \forall i \in \mathcal{S},  \label{eq:admmCon}
	\\
	\sum_{i\in \mathcal{S}} E_i \bar{z}_i &= 0,\label{eq:admmCoupling}
	\end{align}
\end{subequations} where $\gamma_i \in \mathbb{R}^{n_{i}}$ are Lagrange multipliers associated to \eqref{eq:admmCon}, and where
$\mathbb{Z}_i \doteq \{ z_i\in \mathbb{R}^{n_i} \ \vert \ g_i(z_i)=0, h_i(z_i) \leq0 \}$.
Let $z \doteq \text{col}(z_i)_{i \in \mathcal{S}}$, $\bar{z} \doteq \text{col}(\bar{z}_i)_{i \in \mathcal{S}}$, and $\gamma \doteq \text{col}(\gamma_i)_{i \in \mathcal{S}}$ denote the centralized variables in~\eqref{ocp:admm}.
The augmented Lagrangian of \eqref{ocp:admm} is given by
\begin{align*}
L_\rho(z,\bar{z},\gamma) &= \sum_{i\in \mathcal{S}} L_{\rho,i}(z_i, \bar{z}_i,\gamma_i) \\
&=  \sum_{i\in \mathcal{S}} f_i(z_i) + \gamma_i^\top (z_i - \bar{z}_i) + \frac{\rho}{2} \|z_i - \bar{z}_i\|_2^2,
\end{align*} where $\rho \in \mathbb{R}^+$ is called penalty parameter.
The ADMM iterations read
\begin{subequations}
\begin{align}
z_i^{l+1} &= \argmin_{z_i \in \mathbb{Z}_i} L_{\rho,i}(z_i,\bar{z}_i^l,\gamma_i^l),\label{eq:admmZ}\\
\bar{z}^{l+1} &= \argmin_{\bar{z} \in \mathbb{E}} L_{\rho}(z^{l+1},\bar{z},\gamma^l),\label{eq:admmAvg}\\
\gamma_i^{l+1} &= \gamma_i^l + \rho (z_i^{l+1} - \bar{z}_i^{l+1}),\label{eq:admmD}
\end{align}
\end{subequations} where $\mathbb{E} \doteq \{\bar{z} \in \mathbb{R}^n \,|\, E \bar{z} \doteq 0 \},$ and $E \doteq [E_1,\dots, E_S].$
Each subsystem computes the updates~\eqref{eq:admmZ} and~\eqref{eq:admmD} individually.
The update~\eqref{eq:admmAvg} is equivalent to an efficient averaging step if the Lagrange multiplier initialization $\gamma^0$ satisfies
\begin{equation}\label{eq:admmInit}
(I - E^\top (E E^\top)^{-1}E) \gamma^0 = 0.
\end{equation} 
Such an initialization may be obtained by choosing $\gamma^0 = 0$, or by warm-starting with a solution obtained from a previous execution of ADMM.
The averaging procedure is given in Steps~\ref{admm:commz}--\ref{admm:formzbar} of Algorithm~\ref{alg:ADMM}. First, in Step~\ref{admm:commz}, each subsystem sends the updated copy values obtained in Step~\ref{admm:z} to its in-neighbors and receives the updated copies corresponding to its own trajectory from its out-neighbors. 
In Step~\ref{admm:avg}, each subsystem averages all decision variables corresponding to its state trajectory.
Then, in Step~\ref{admm:commzbar}, each subsystem shares the averaged values with all out-neighbors and receives the averaged values from all in-neighbors.
Finally, in Step~\ref{admm:formzbar}, each subsystem forms $\bar{z}_i^{l+1}$ from the mean values.
Upon ADMM convergence,~\eqref{eq:admmCon} holds and the subsystems have reached consensus.

ADMM has successfully been used both for DMPC of linear systems~\citep{Rostami2017} and for nonlinear DMPC~\citep{Bestler2019}.
If OCP~\eqref{ocp} is an NLP, then each robot solves an NLP in Step~\ref{admm:z} of ADMM.
This is computationally expensive and using general-purpose NLP solvers such as \texttt{IPOPT}~\citep{Wachter2006}--especially if deployed on embedded hardware---may limit the number of ADMM iterations per MPC step~\citep{vanParys2017}.
If instead OCP~\eqref{ocp} is a QP, then each robot only solves a QP in each ADMM iteration, which is computationally much less demanding.

ADMM is guaranteed to converge for convex problems~\citep{Boyd2011,He2012}. 
While ADMM is guaranteed to converge also for some non-convex problems~\citep{Wang2019}, there exist no generally applicable convergence results if OCP~\eqref{ocp} is non-convex as is the case in distributed nonlinear MPC.
We therefore use Algorithm~\ref{alg:ADMM} if OCP~\eqref{ocp}, or respectively its reformulation~\eqref{nlp}, is a convex QP.

\begin{algorithm}[t]
	\caption{\hspace*{-0.02cm}ADMM for OCP~\eqref{nlp}~\citep{Boyd2011}} \label{alg:ADMM}
	\begin{algorithmic}[1]
		\State Initialization: $l = 0$, $\bar{z}_i^0,\gamma_{i}^0$ satisfying~\eqref{eq:admmInit} for all $i \in \mathcal{S}$, $l_\mathrm{max}$
		\While{$l < l_\mathrm{max}$ for all $i \in \mathcal{S}$}
		\State Solve subsystem program $
		z_i^{l+1} = \displaystyle\argmin_{z_i \in \mathbb{Z}_i} L_{\rho,i}(z_i,\bar{z}_i^l,\gamma_i^l)
		$\label{admm:z}
		\State Receive copies $\boldsymbol{w}^{l+1}_{ij}$ from all ${j \in \mathcal{N}_i^{\text{out}}}$\label{admm:commz}
		\State Average received copies with original trajectories
		\begin{equation*}
		\bar{\boldsymbol{x}}^{l+1}_i =  \frac{1}{\lvert \mathcal{N}_i^{\text{out}} \rvert +1} \left(\boldsymbol{x}_i^{l+1} + \sum_{j \in \mathcal{N}_i^{\text{out}}} \boldsymbol{w}^{l+1}_{ij} \right)
		\end{equation*}\label{admm:avg}  
		\State Send average $\bar{\boldsymbol{x}}^{l+1}_i$ to all $j \in \mathcal{N}_i^{\text{out}}$\label{admm:commzbar}
		\State Form $\bar{z}_i^{l+1}=\text{col}(\bar{\boldsymbol{x}}^{l+1}_i,\bar{\boldsymbol{u}}^{l+1}_i,\bar{\boldsymbol{w}}^{l+1}_{i}), \, \bar{\boldsymbol{w}}^{l+1}_i = \text{col}(\bar{\boldsymbol{x}}^{l+1}_j)_{j \in \mathcal{N}_i^\text{in}}
		$\label{admm:formzbar}
		\State Update dual variable $\gamma_{i}^{l+1} =\gamma_{i}^l + \rho (z_i^{l+1}-\bar{z}_i^{l+1})$
		\State $l = l+1$
		\EndWhile
		\State \textbf{return} $z_i^l$
	\end{algorithmic}
\end{algorithm}

\subsection{The Decentralized SQP Method for Nonlinear DMPC}
To solve non-convex OCPs, we consider the dSQP algorithm proposed by~\citep{Stomberg2022a}.
Similar to~\citep{Engelmann2020c}, the method admits a bi-level structure, i.e., we index outer SQP iterations by $q$ and inner ADMM iterations by~$l$. 
Let $\nu \doteq \text{col}(\nu_i)_{i \in \mathcal{S}}$ and $\mu \doteq \text{col}(\mu_i)_{i \in \mathcal{S}}$ denote the centralized dual variables of problem~\eqref{nlp}.
Given a primal-dual iterate~$(z^q,\nu^q,\mu^q,\lambda^q)$, we approximate~\eqref{nlp} by
\begin{subequations}\label{eq:qpk}
	\begin{align} 
	\min_{\Delta z_i \in \mathbb{R}^{n_i}, i \in \mathcal{S}} \; \sum_{i \in \mathcal{S}} \frac{1}{2} \Delta z_i^\top H_{i}^q \Delta z_i &+ \nabla f_i(z_i^q)^\top \Delta z_i \\
	\nonumber\text{subject to}\hspace{1.8cm}&\\
    g_i(z_i^q) + \nabla g_i(z_i^q)^\top \Delta z_i&=0 \quad \; \vert\; \nu_i^{\text{QP}}, \quad \forall i \in \mathcal{S},\\
	h_i(z_i^q) + \nabla h_i(z_i^q)^\top \Delta z_i &\leq 0 \quad \; \vert\; \mu_i^{\text{QP}}, \quad \forall i \in \mathcal{S},\\
	\sum_{i \in \mathcal{S}} E_i(z_i^q + \Delta z_i) &= 0 \quad \; \vert\; \lambda^{\text{QP}},
	\end{align}
\end{subequations} 
where the matrix $H_{i}^q \approx  \nabla_{z_iz_i}^2 L_i^q \doteq \nabla_{z_iz_i}^2 L_i(z_i^q,\nu_i^q,\mu_i^q,\lambda^q)$ is positive definite, and where the Lagrangian function of subsystem $i$ is given by ${L_i(\cdot) = f_i(z_i) + \nu_i^\top g_i(z_i) + \mu_i^\top h_i(z_i) + \lambda^\top E_i z_i}.$ 
The convex quadratic approximation~\eqref{eq:qpk} is solved in decentralized fashion via ADMM to obtain the SQP steps ${\Delta z_i = \text{col}(\Delta \boldsymbol{x}_i, \Delta \boldsymbol{u}_i, \Delta \boldsymbol{w}_i)}$, ${\Delta \nu_i \doteq \nu_i^{\text{QP}} - \nu_i^q}$, $\Delta \mu_i \doteq \mu_i^{\text{QP}} - \mu_i^q$, and $\Delta \lambda \doteq \lambda^{\text{QP}} - \lambda^q$.

Algorithm~\ref{alg:dSQP} summarizes the dSQP iterations. For the sake of compact notation, we collect the subsystem constraint sets of QP~\eqref{eq:qpk} in
\begin{align*}
\mathbb{S}_i^q \doteq \left\{ \Delta z_i \in \mathbb{R}^{n_i} \,\left|\, \begin{aligned} g_i(z_i^q) + \nabla g_i(z_i^q)^\top \Delta z_i &= 0\\
h_i(z_i^q) + \nabla h_i(z_i^q)^\top \Delta z_i &\leq 0 \end{aligned} \right. \right\}
\end{align*} and the augmented Lagrangian function of subsystem~$i$ is
\begin{equation*}
L_{\rho,i}^{QP,q} = \frac{1}{2} \Delta z_i^\top H_{i}^q \Delta z_i + \nabla f_i(z_i^q)^\top \Delta z_i + \gamma_i^{\top} (\Delta z_i - \Delta \bar{z}_i) + \frac{\rho}{2} \| \Delta z_i - \Delta \bar{z}_i\|_2^2.
\end{equation*}

\begin{rem}[Local dSQP convergence]\label{rem:dsqpConv}
	
	We terminate ADMM after $l_\mathrm{max}$ iterations per outer iteration in Step~\ref{dsqp:while} of Algorithm~\ref{alg:dSQP}.
	This is due to practical reasons and limits the execution time of dSQP in each MPC step.	
	However, one could also terminate ADMM in Step~\ref{dsqp:while} of Algorithm~\ref{alg:dSQP} dynamically to obtain local convergence guarantees for dSQP.
	To obtain the convergence result, we define ${d_i \doteq \text{col}(\Delta z_i,\Delta \nu_i,\Delta \mu_i)}$, $d = ( \text{col}(d_i)_{i \in \mathcal{S}},\Delta \lambda)$, and ${p\doteq\text{col}(z,\nu,\mu,\lambda)}$.
	Let ${p^\star\doteq\text{col}(z^\star,\nu^\star,\mu^\star,\lambda^\star)}$ be a Karush-Kuhn-Tucker (KKT) point of~\eqref{nlp}. 
	We denote the set of active inequality constraints at $z_i^\star$ by $\mathcal{A}_i \doteq \{j \in \{1,\dots,n_{h,i} \}\ \vert\ [h_i(x_i^\star)]_j = 0\}$. 
	The stopping criterion reads	
	\begin{equation}\label{eq:modStop}
	\| \tilde{F}^q + \nabla \tilde{F}^q d \| \leq \eta^q \| \tilde{F}^q \|,
	\end{equation} 
	where
	\begin{align*}
	\tilde{F} \doteq \begin{bmatrix}
	\nabla_{z_1} L_1(z_1,\nu_1,\mu_1,\lambda)\\
	g_1(z_1)\vspace*{-0.1cm}\\
	\vdots\vspace*{-0.1cm}\\
	\nabla_{z_S} L_S(z_S,\nu_S,\mu_S,\lambda)\\
	g_S(z_S)\\
	\sum_{i \in \mathcal{S}}E_i z_i
	\end{bmatrix},
	\nabla \tilde{F} \hspace*{-0.1cm}= \hspace*{-0.1cm}\begin{bmatrix} 
	\nabla \tilde{F}_1 & \hspace*{-0.3cm} \dots  &\hspace*{-0.3cm} 0 & \hspace*{-0.3cm}\bar{E}_1^\top\vspace*{-0.1cm}\\
	\vdots & \hspace*{-0.3cm}\ddots & \hspace*{-0.3cm}\vdots & \vdots\\
	0 & \hspace*{-0.3cm}\dots & \hspace*{-0.3cm}\nabla \tilde{F}_S & \hspace*{-0.3cm}\bar{E}_S^\top\\
	\tilde{E}_1 & \hspace*{-0.3cm}\dots & \hspace*{-0.3cm}\tilde{E}_S & \hspace*{-0.3cm}0
	\end{bmatrix},\\	
	\nabla \tilde{F}_i \doteq  \begin{bmatrix}
	\nabla_{z_iz_i}^2 L_i & \nabla g_i(z_i) & \nabla h_i(z_i)\\
	\nabla g_i(z_i)^\top & 0 & 0 \\
	\end{bmatrix},
	\end{align*} with $\tilde{E}_i \doteq \begin{bmatrix} E_i & 0 & 0 \end{bmatrix}$, $\bar{E}_i \doteq \begin{bmatrix} E_i & 0\end{bmatrix}$, and $\eta^q \geq 0$.	
	If this stopping criterion is employed, then dSQP is guaranteed to converge locally as described in Theorem~\ref{thm:dsqpConv} below, which appeared as Theorem~2 in \citep{Stomberg2022a}. \hfill $\square$
\end{rem}

\begin{thm}[dSQP convergence \citep{Stomberg2022a}]\label{thm:dsqpConv}
	Let $p^\star$ be a KKT point of \eqref{nlp} which, for all $i \in \mathcal{S}$, satisfies
	\begin{enumerate}
		\item[i)] $h_i(z_i^\star) + \mu_i^\star \neq 0$ (strict complementarity), 
		\item[ii)] $\Delta z_i^\top \nabla_{z_iz_i}^2 L_i(z_i^\star,\nu_i^\star,\mu_i^\star,\lambda^\star)\Delta z_i > 0$ for all $\Delta  z_i \neq 0$ with $\nabla g_i(x_i^\star)^\top \Delta z_i = 0$.
	\end{enumerate}
	Furthermore, suppose the matrix
	\begin{align*}
	\begin{bmatrix} \nabla_{z_1} g_1(z_1^\star)^\top & &\vspace*{-0.2cm}\\
	&\ddots & \vspace*{-0.2cm}\\
	& & \nabla_{z_S} g_S(z_S^\star)^\top\\
	\nabla_{z_1} [h_1(z_1^\star)]_{\mathcal{A}_1}^\top & & \vspace*{-0.2cm}\\
	& \ddots & \vspace*{-0.2cm}\\
	& & \nabla_{z_S} [h_S(z_S^\star)]_{\mathcal{A}_S}^\top\\
	E_1 & \dots & E_S \end{bmatrix}
	\end{align*} has full row rank, i.e., $z^\star$ satisfies the linear independence constraint qualification.	
	Furthermore, suppose the Hessian approximation ${H_i^q =  \nabla_{z_iz_i}^2 L_i^q}$ is used and that ADMM is terminated in~Step~\ref{dsqp:while} of Algorithm~\ref{alg:dSQP} dynamically based on~\eqref{eq:modStop}.
	Then, there exist $\varepsilon > 0$ and $\eta > 0$ such that for all $p^0 \in \mathcal{B}_{\varepsilon} ( p^\star)$ the following holds:
	\begin{enumerate}
		\item[i)] If $\eta^q \leq \eta$, then the sequence $\{p^q\}$ generated by Algorithm~\ref{alg:dSQP} converges to~$p^\star$ and the convergence rate is q-linear in the outer iterations.
		\item[ii)] If $\eta^q \rightarrow 0$, then the convergence rate is q-superlinear in the outer iterations.
		\item[iii)] If $\eta^q =O(\| \tilde{F}^q \|)$  and if
		${\nabla_{z_iz_i}^2 f_i,\nabla_{z_iz_i}^2 [g_i]_j,j \in \{1,\dots,n_{g,i}\}}$, and ${\nabla_{z_iz_i}^2 [h_i]_j, j \in \{1,\dots,n_{h,i}\}}$ are Lipschitz continuous inside $\mathcal{B}_{\varepsilon}$, then the convergence rate is q-quadratic in the outer iterations. \hfill$\square$
	\end{enumerate}
\end{thm}

\begin{rem}[Hessian approximation]	 
	There exist two common choices for~$H_i^q$.
	The first option is to set ${H_i^q = \nabla_{z_iz_i}^2 L_i^q}$, and to regularize if needed.
	For the regularization, we follow a heuristic by changing the sign of negative eigenvalues~\citep{Engelmann2020f}. That is, we first compute an eigenvalue decomposition $\nabla_{z_iz_i}^2 L_i^q = V_i \Lambda_i V_i^\top$. Then, we set $H_i^q = V_i \tilde{\Lambda}_i V_i^\top$, where for all $j \in \{1,\dots,n_i\}$
	\begin{equation*}
	\tilde{\Lambda}_{i,jj} \doteq \left\{ \begin{array}{ll} \varepsilon, &\text{if } \Lambda_{i,jj} \in [-\varepsilon,\varepsilon],\\
	|\Lambda_{jj} |, &\text{else},
	\end{array}\right.
	\end{equation*} and $\varepsilon = 10^{-4}$. 
	The second option is the Gauss-Newton approximation $H_i^q = \nabla^2_{z_iz_i} f_i(z_i)$~\citep{Bock1983}.
	In this case, $H_i^q$ is constant, strongly convex, and can be evaluated offline, because the objectives $f_i(\cdot)$ are strongly convex quadratic functions. \hfill $\square$	
\end{rem}

\begin{algorithm}[t]
	\caption{dSQP for OCP~\eqref{nlp}~\citep{Stomberg2022a}} \label{alg:dSQP}
	\begin{algorithmic}[1]
		\State Initialization: $q=0,z_i^0,\nu_i^0,\mu_i^0,\gamma_i^0$ satisfying~\eqref{eq:admmInit} for all $i\in \mathcal{S},q_\mathrm{max},l_\mathrm{max}$
		\While{$q < q_\mathrm{max}$ for all $i \in \mathcal{S}$}
		\State Compute $\nabla f_i(z_i^q), \nabla g_i(z_i^q), \nabla h_i(z_i^q), H_i^q$ locally\label{dsqp:H}
		\State ADMM initialization $l=0, \Delta\bar{z}_i^0 = 0$
		\While{$l < l_\mathrm{max}$}\label{dsqp:while}
				\State Solve QP $
				\Delta z_i^{l+1} = \displaystyle\argmin_{\Delta z_i \in \mathbb{S}_i}  L_{\rho,i}^{QP,q}(\Delta z_i, \Delta \bar{z}_i^l, \gamma_i^l)$\label{dsqp:z}
		\State Receive copies $\Delta \boldsymbol{w}^{l+1}_{ij}$ from all ${j \in \mathcal{N}_i^{\text{out}}}$
		\State Average received copies with original trajectories
		\begin{equation*}
		\Delta \bar{\boldsymbol{x}}^{l+1}_i =  \frac{1}{\lvert \mathcal{N}_i^{\text{out}} \rvert +1} \left(\Delta \boldsymbol{x}_i^{l+1} + \sum_{j \in \mathcal{N}_i^{\text{out}}} \Delta \boldsymbol{w}^{l+1}_{ij} \right)
		\end{equation*}
		\State Send average $\Delta \bar{\boldsymbol{x}}^{l+1}_i$ to all $j \in \mathcal{N}_i^{\text{out}}$
		\State $\Delta \bar{z}_i^{l+1}=\text{col}(\Delta \bar{\boldsymbol{x}}^{l+1}_i,\Delta \bar{\boldsymbol{u}}^{l+1}_i,\Delta \bar{\boldsymbol{w}}^{l+1}_{i})$,
		 
		\hspace*{0.33cm}$\Delta \bar{\boldsymbol{w}}^{l+1}_i = \text{col}(\Delta \bar{\boldsymbol{x}}^{l+1}_j)_{j \in \mathcal{N}_i^\text{in}}$
		\State Update dual variable $\gamma_{i}^{l+1} =\gamma_{i}^l + \rho (\Delta z_i^{l+1}-\Delta \bar{z}_i^{l+1})$
		\State $l = l+1$
		\EndWhile
		\State $z_i^{q+1} \hspace*{-0.1cm}=\hspace*{-0.1cm} z_i^q + \Delta \bar{z}_i^l, \nu_i^{q+1} \hspace*{-0.1cm}=\hspace*{-0.1cm} \nu_i^{l,QP}, \mu_i^{q+1} \hspace*{-0.1cm}=\hspace*{-0.1cm} \mu_i^{l,QP}, \gamma_i^{q+1} \hspace*{-0.1cm}=\hspace*{-0.1cm} \gamma_i^l$
		\State $q = q+1$
		\EndWhile
		\State \textbf{return} $z_i^q$
	\end{algorithmic}
\end{algorithm} 

We conclude this section by summarizing the main characteristics of ADMM and dSQP if OCP~\eqref{ocp} is either a convex QP or a non-convex NLP.
Under Assumption~\ref{ass:ocp}, OCP~\eqref{ocp} is a convex QP if the dynamics are linear and if the system constraints are convex polytopes.
In this case ADMM and dSQP are equivalent, require each subsystem to solve one convex QP per iteration, and are guaranteed to converge to a global optimum~\citep{Boyd2011}.
If the dynamics are non-linear, or if the system constraints are non-convex, or both, then OCP~\eqref{ocp} is a non-convex NLP. 
In this case, there exist no convergence guarantees for ADMM and ADMM requires each subsystem to solve an NLP in every iteration.
However, there exist nominal local convergence guarantees for dSQP, cf. Remark~\ref{rem:dsqpConv} and~\citep{Stomberg2022a}. 
In addition, dSQP only requires each subsystem to solve a convex QP per inner iteration.
Table~\ref{tab:optAlg} summarizes these key properties and we refer to~\citep{Stomberg2022} for an in-depth overview of further decentralized methods for linear-quadratic DMPC.

\begin{table}
    \caption{ADMM and dSQP properties.}
    \label{tab:optAlg}
	\centering
	\scalebox{0.95}{
	\begin{tabular}{l l l l}
		Method & Centralized & Problem solved & Convergence\\
			&	problem type	& by subsystem & guarantees\\
		\hline
		ADMM & convex QP & convex QP & yes\vspace{0.2cm}\\
		dSQP & convex QP & convex QP & yes\vspace{0.2cm}\\
		\multirow{2}{4pt}{ADMM} & non-convex & non-convex & \multirow{2}{4pt}{no}\\
			& NLP & NLP & \vspace{0.2cm}\\ 
		\multirow{2}{4pt}{dSQP} & non-convex & convex & yes, see\\
		& NLP & QP & Remark~\ref{rem:dsqpConv}\\ 
	\end{tabular}
}
\end{table}

\section{Implementation and Hardware}
\label{sec:implementation}
\subsection{Hardware and Experiment Environment}
The proposed control algorithms are tested in a hardware environment designed for distributed robotics and networked computation. 
The setup exhibits the full set of challenges arising in distributed control, i.e., plant-model mismatch, lossy and delayed communication, and networked computation.

The robots are controlled using a bi-level control architecture with DMPC on the upper level and PID controllers on the lower level.
The DMPC controller computes the desired velocity with respect to an inertial frame of reference.
The PID controller then commands the motors to attain the angular wheel velocities which, ideally, under the assumption of rolling without slipping, kinematically correspond to the desired velocity of the robot body.
The sampling rate is $5\,\text{Hz}$ for the DMPC controllers and $100\,\text{Hz}$ for the PID controllers. 
	
Since real, imperfect, hardware is used, there is a non-negligible plant-model mismatch for the DMPC controller.
Indeed, although built in the same manner, each individual robot behaves slightly differently than the other robots~\citep{EschmannEbelEberhard21b}, i.e., the robots exhibit characteristic nonlinear behavior, due to friction, the complicated wheel-floor contacts, and manufacturing imperfections.

\begin{figure}
	\includegraphics[width=\columnwidth]{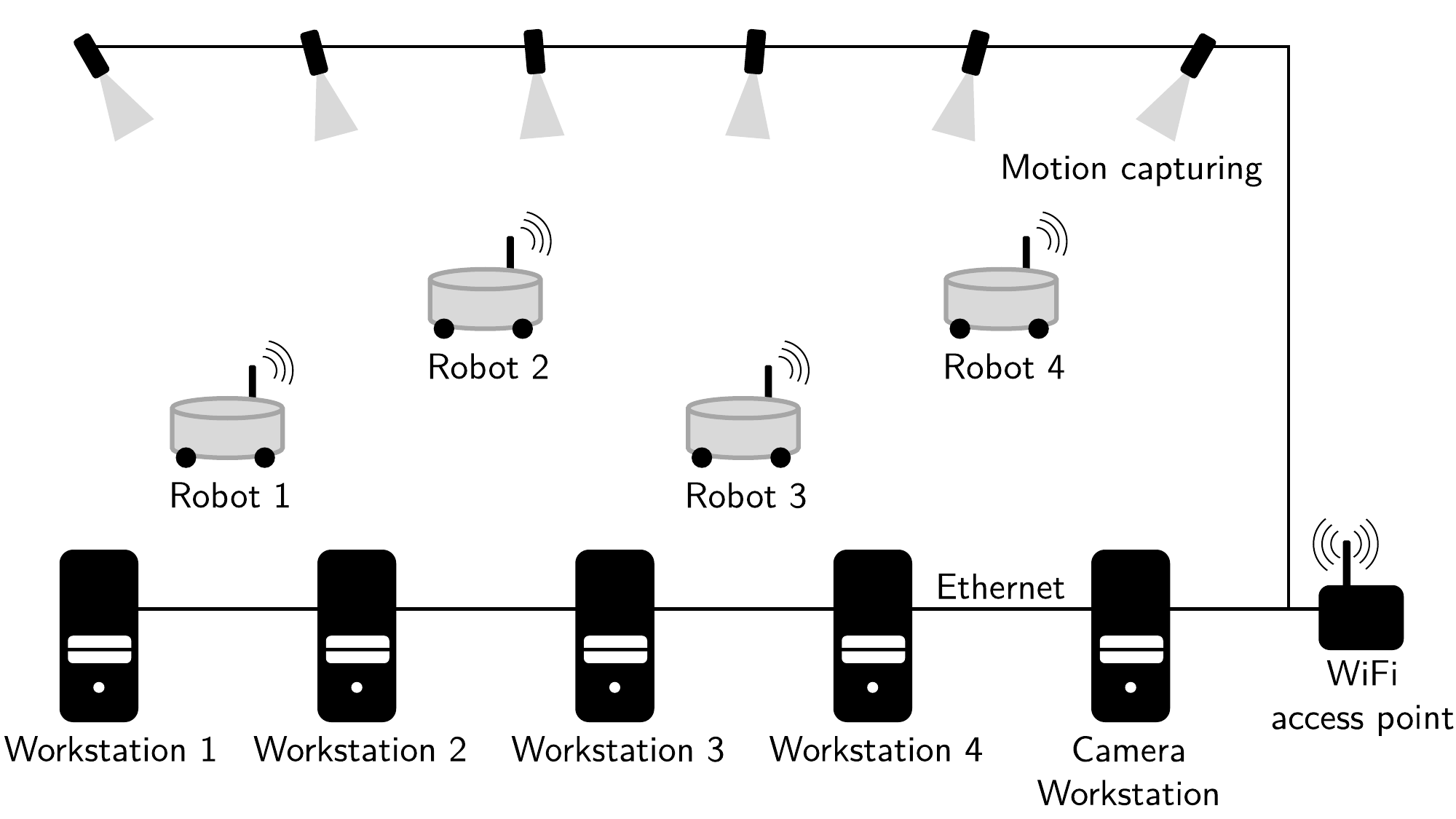}
	\caption{Experimental setup. The DMPC controllers run on the workstation computers and each robot is assigned one workstation to run Algorithm~\ref{alg:DMPC}.}\label{fig:setup}
\end{figure}

To arrive at a realistic networked setting, we assign a dedicated workstation computer to each robot to run the DMPC controller as shown in Figure~\ref{fig:setup}. 
However, these computers are not mounted on the robots to keep the robot design simple and lightweight.
The PID controllers run on microcontrollers onboard the robots. 
The DMPC control input is sent wirelessly to the robots via User Datagram Protocol (UDP) multicast and the \texttt{LCM} library ~\citep{HuangOlsonMoore10}.  
The robot poses are tracked by an Optitrack tracking system consisting of six Prime 13W cameras.
The cameras use the commercial Motive software, which runs on a separate computer. 
A custom software republishes the robot pose data into messages delivered to the computers running the DMPC controller via UDP multicast  and \texttt{LCM}. 

Moreover, the computing environment is heterogeneous as the computers assigned to the robots differ in their processing capability.
This increases the chance that the programs need to wait for each other to complete the calculations in the DMPC controller. 
Specifically, the computers are equipped with either an Intel Core i7-9750H ($2.6\, \textnormal{GHz}$), an Intel Xeon E5530 ($2.4\, \textnormal{GHz}$), or an Intel Xeon E31245 ($3.3\, \textnormal{GHz}$) processor and they run Debian linux. 

\subsection{DMPC Implementation}

We have implemented Algorithms~\ref{alg:DMPC}--\ref{alg:dSQP} in C++ and the source code is available online\footnote{\url{https://github.com/OptCon/dmpc_rto}}.
Besides the C++ standard library, our implementation relies on four external libraries: (i) \texttt{Eigen}\footnote{\url{https://eigen.tuxfamily.org}} to store vectors and matrices as well as to regularize $H_i$; (ii) \texttt{qpOASES}~\citep{Ferreau2014} to solve the subsystem QPs; (iii) \texttt{CasADi}~\citep{Andersson2019} to evaluate sensitivities in dSQP via automatic differentiation; and (iv) \texttt{LCM} to exchange optimizer iterates with neighbors.

While our implementation uses dynamic memory, we allocate any dynamic memory prior to running the control loop, thus accelerating the execution time of our methods.

For Step~\ref{admm:z} in Algorithm~\ref{alg:ADMM} and for Step~\ref{dsqp:z} in Algorithm~\ref{alg:dSQP}, we use the online active-set method implemented in \texttt{qpOASES}.
We initialize the QP solver once prior to running the DMPC controller, and solely use the efficient hotstart procedure online to update the current and desired robot positions.

All communication of the algorithms happens via the aforementioned \texttt{LCM} library. 
There, communication follows a publish-subscribe pattern. 
Messages are received and cached asynchronously by a separate thread of the solver program, with the data being accessed when required by the algorithm. 
If a required piece of data has not yet been received and cached, the algorithm waits until all data necessary to continue calculations is available. 
For state measurements to be received, a maximum waiting time can be set so that the algorithm does not wait indefinitely if a state measurement has been lost.
For optimizer iterates, however, lost messages are resent by the control application to ensure a synchronous execution of ADMM.
This is necessary since we rely on UDP and not on the transmission control protocol as~\citep{Burk2021}. 

The main thread of each DMPC program sleeps after Step~6 of Algorithm~\ref{alg:DMPC}.
To reduce the effect of jitter, we implement the sleep via \texttt{std::this\_thread::sleep\_until}.
This is crucial to ensure that all controllers keep the same sampling interval $\Delta t$ in the distributed setting.

\section{Experimental Results}\label{sec:exp}

\subsection{Formation Control via linear-quadratic MPC and ADMM}

\begin{figure}[b]
	\includegraphics[width=\columnwidth]{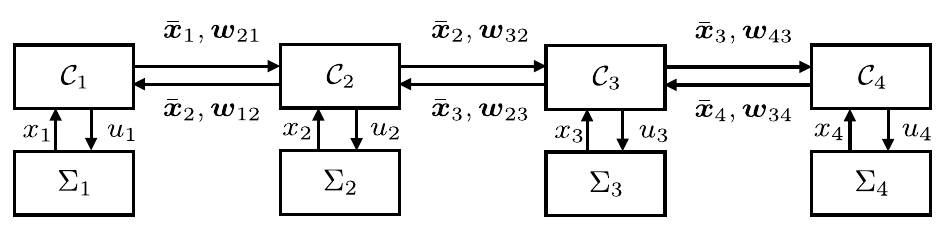}
	\caption{Block diagram for a path graph of four robots $\Sigma_i$. The controllers $\mathcal{C}_i$ of neighboring robots exchange predicted position trajectories $\bar{\boldsymbol{x}}_i$ and $\boldsymbol{w}_{ij}$.}\label{fig:blockDiagram}
\end{figure}

We consider a scenario where a set $\mathcal{S} = \{1,2,3,4\}$ of four mobile robots moves in a chain-like formation. 
OCP~\eqref{ocp} is designed such that each robot is coupled to its neighbors in the chain, i.e., the underlying coupling and communication graph is a path graph of length four.
The robots are initially positioned with an equidistant spacing of $0.4\,\mathrm{m}$. 
The first robot then moves along a rectangular path and the remaining robots follow while keeping the initialized distance to their neighbors.
Figure~\ref{fig:blockDiagram} summarizes the control architecture and shows the communicated signals between robots and controllers.

We choose a horizon $N=7$ and a sampling interval ${\Delta t = 0.2\,\text{s}}$.
For the cost matrices, we set $Q_{11} = Q_{22} = Q_{33} = \text{diag}(20,20)$, $Q_{44} = \text{diag}(10,10)$, $Q_{ij} = \text{diag}(-10,-10)$ for all $j \in \{i-1,i+1\} \cap \mathcal{S} $ and for all $i \in \mathcal{S}$, and $R_{ii} = \text{diag}(1,1)$ for all $i \in \mathcal{S}$. 
The remaining cost matrices $Q_{ij}$ are set to zero. 
The terminal cost matrix is chosen as $P = Q$. The setpoints are $\bar{x}_1 = x_1^{\text{d}}(t)$ and $\bar{x}_i = \bar{x}_{i-1} + \delta$ for all $i \in \{2,3,4\}$, where $x_1^{\text{d}}(t)$ encodes the rectangular path and ${\delta = \begin{bmatrix}-0.4 & 0\end{bmatrix}^\top\,\mathrm{m}}$ is the desired spacing between neighbors.
The input constraints are given as ${\begin{bmatrix} -0.2 & -0.2 \end{bmatrix}^\top\,\mathrm{m/s} \leq u_i \leq \begin{bmatrix} 0.2 & 0.2 \end{bmatrix}}^\top \,\mathrm{m/s}$ for all $i \in \mathcal{S}$.
We enforce no state constraints such that $\mathbb{X}_i = \mathbb{R}^{n_{x,i}}$ and $\mathbb{X}_{ij} = \mathbb{R}^{n_{x,i}} \times \mathbb{R}^{n_{x,j}}$ for all $i,j \in \mathcal{S}$.
OCP~\eqref{ocp} therefore is a strongly convex QP.

\begin{figure}[t]
	\def\svgwidth{0.4977\columnwidth}%
\begingroup%
  \makeatletter%
  \providecommand\color[2][]{%
    \errmessage{(Inkscape) Color is used for the text in Inkscape, but the package 'color.sty' is not loaded}%
    \renewcommand\color[2][]{}%
  }%
  \providecommand\transparent[1]{%
    \errmessage{(Inkscape) Transparency is used (non-zero) for the text in Inkscape, but the package 'transparent.sty' is not loaded}%
    \renewcommand\transparent[1]{}%
  }%
  \providecommand\rotatebox[2]{#2}%
  \newcommand*\fsize{\dimexpr\f@size pt\relax}%
  \newcommand*\lineheight[1]{\fontsize{\fsize}{#1\fsize}\selectfont}%
  \ifx\svgwidth\undefined%
    \setlength{\unitlength}{1078.25033569bp}%
    \ifx\svgscale\undefined%
      \relax%
    \else%
      \setlength{\unitlength}{\unitlength * \real{\svgscale}}%
    \fi%
  \else%
    \setlength{\unitlength}{\svgwidth}%
  \fi%
  \global\let\svgwidth\undefined%
  \global\let\svgscale\undefined%
  \makeatother%
  \begin{picture}(1,0.75028964)%
    \lineheight{1}%
    \setlength\tabcolsep{0pt}%
    \put(0,0){\includegraphics[width=\unitlength,page=1]{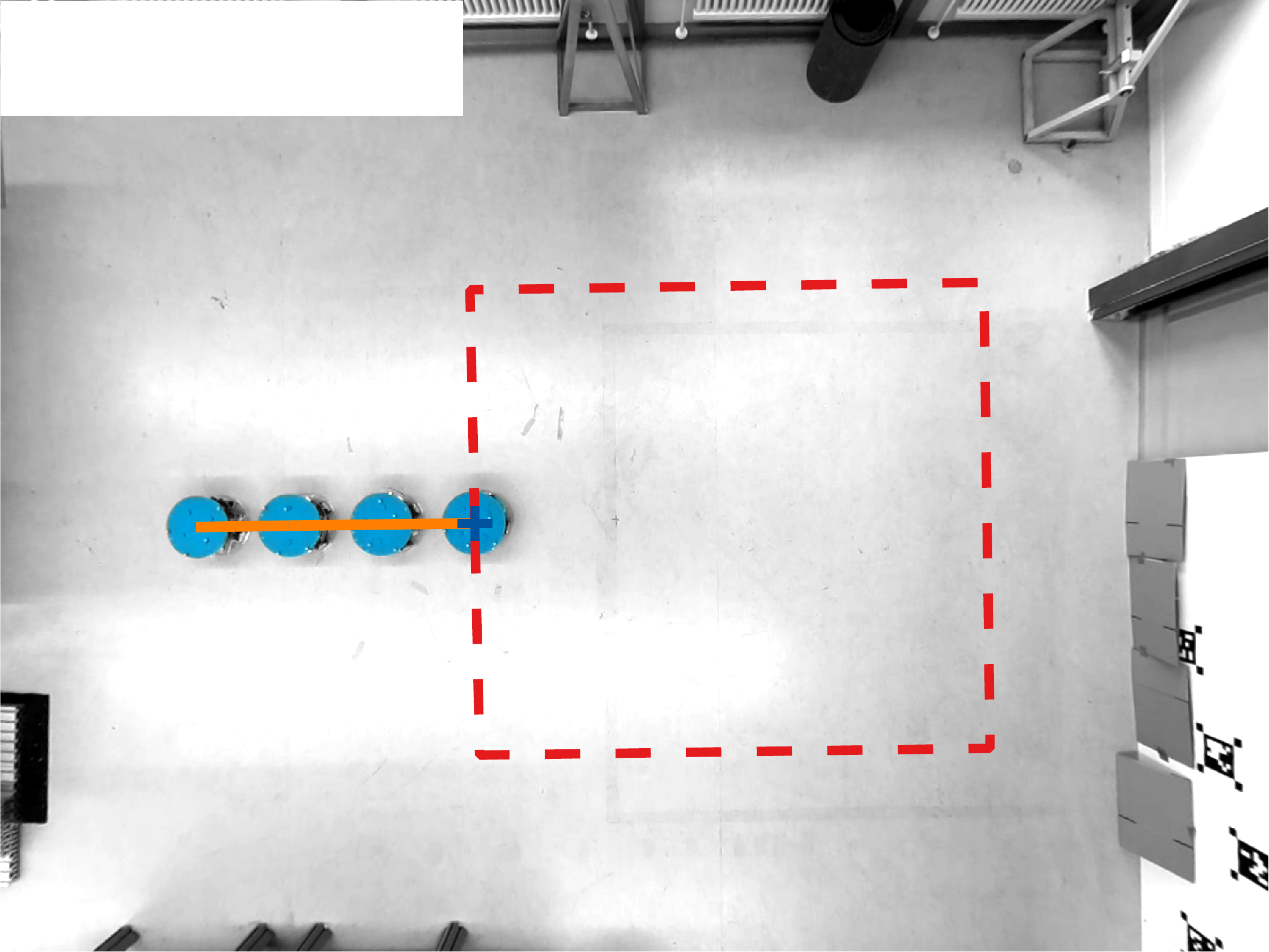}}%
    \put(0.0118159,0.67814321){\makebox(0,0)[lt]{\lineheight{1.25}\smash{\begin{tabular}[t]{l}$t = \phantom{0}0.00\,\textnormal{s}$\end{tabular}}}}%
    \put(0,0){\includegraphics[width=\unitlength,page=2]{exp_ADMM_1.pdf}}%
  \end{picture}%
\endgroup%
\hfill%
	\def\svgwidth{0.4977\columnwidth}%
\begingroup%
  \makeatletter%
  \providecommand\color[2][]{%
    \errmessage{(Inkscape) Color is used for the text in Inkscape, but the package 'color.sty' is not loaded}%
    \renewcommand\color[2][]{}%
  }%
  \providecommand\transparent[1]{%
    \errmessage{(Inkscape) Transparency is used (non-zero) for the text in Inkscape, but the package 'transparent.sty' is not loaded}%
    \renewcommand\transparent[1]{}%
  }%
  \providecommand\rotatebox[2]{#2}%
  \newcommand*\fsize{\dimexpr\f@size pt\relax}%
  \newcommand*\lineheight[1]{\fontsize{\fsize}{#1\fsize}\selectfont}%
  \ifx\svgwidth\undefined%
    \setlength{\unitlength}{1078.25033569bp}%
    \ifx\svgscale\undefined%
      \relax%
    \else%
      \setlength{\unitlength}{\unitlength * \real{\svgscale}}%
    \fi%
  \else%
    \setlength{\unitlength}{\svgwidth}%
  \fi%
  \global\let\svgwidth\undefined%
  \global\let\svgscale\undefined%
  \makeatother%
  \begin{picture}(1,0.75028964)%
    \lineheight{1}%
    \setlength\tabcolsep{0pt}%
    \put(0,0){\includegraphics[width=\unitlength,page=1]{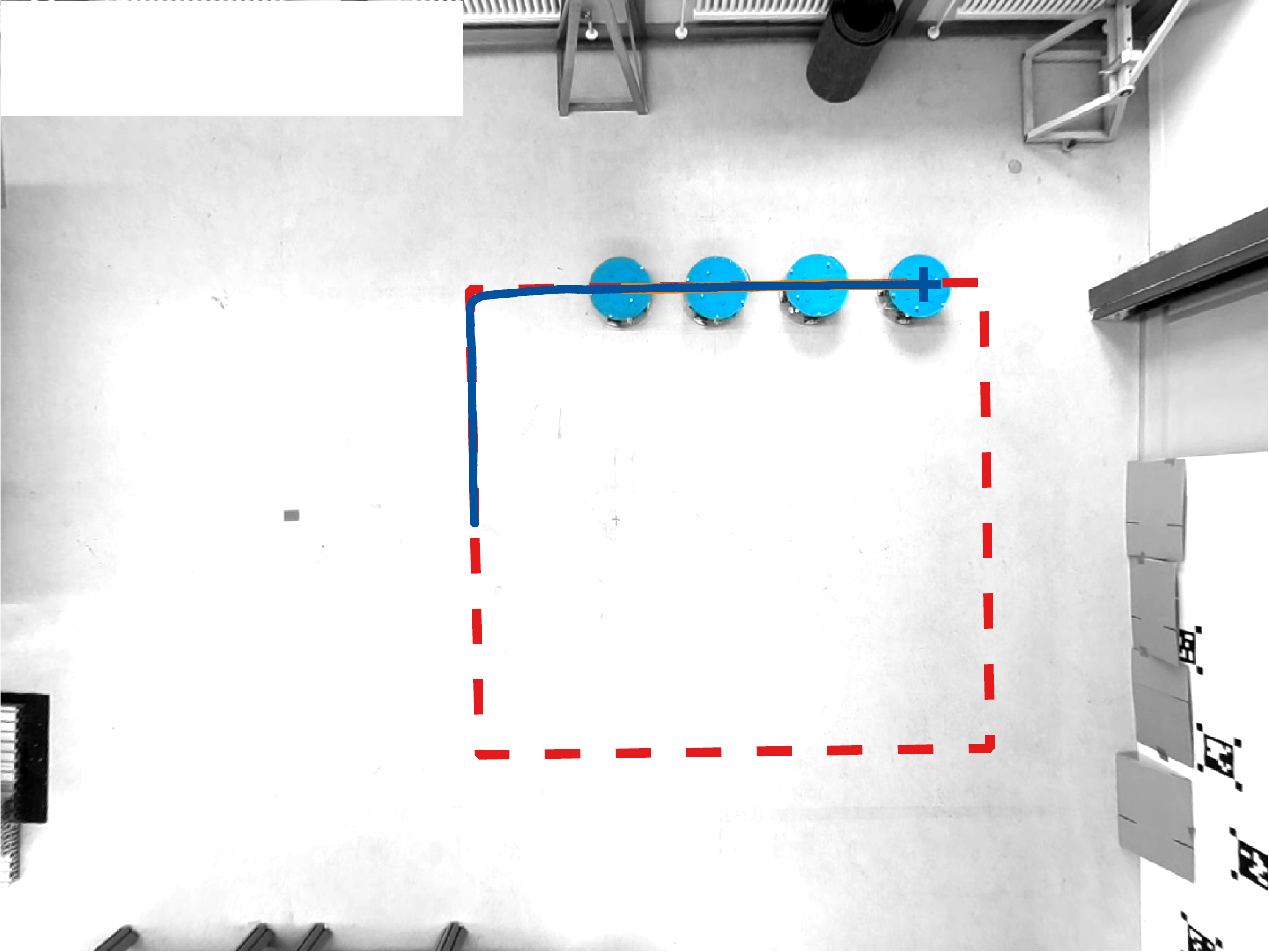}}%
    \put(0.0118159,0.67814321){\makebox(0,0)[lt]{\lineheight{1.25}\smash{\begin{tabular}[t]{l}$t = 35.00\,\textnormal{s}$\end{tabular}}}}%
    \put(0,0){\includegraphics[width=\unitlength,page=2]{exp_ADMM_2.pdf}}%
  \end{picture}%
\endgroup%
\\%
	\def\svgwidth{0.4977\columnwidth}%
\begingroup%
  \makeatletter%
  \providecommand\color[2][]{%
    \errmessage{(Inkscape) Color is used for the text in Inkscape, but the package 'color.sty' is not loaded}%
    \renewcommand\color[2][]{}%
  }%
  \providecommand\transparent[1]{%
    \errmessage{(Inkscape) Transparency is used (non-zero) for the text in Inkscape, but the package 'transparent.sty' is not loaded}%
    \renewcommand\transparent[1]{}%
  }%
  \providecommand\rotatebox[2]{#2}%
  \newcommand*\fsize{\dimexpr\f@size pt\relax}%
  \newcommand*\lineheight[1]{\fontsize{\fsize}{#1\fsize}\selectfont}%
  \ifx\svgwidth\undefined%
    \setlength{\unitlength}{1078.25033569bp}%
    \ifx\svgscale\undefined%
      \relax%
    \else%
      \setlength{\unitlength}{\unitlength * \real{\svgscale}}%
    \fi%
  \else%
    \setlength{\unitlength}{\svgwidth}%
  \fi%
  \global\let\svgwidth\undefined%
  \global\let\svgscale\undefined%
  \makeatother%
  \begin{picture}(1,0.75028964)%
    \lineheight{1}%
    \setlength\tabcolsep{0pt}%
    \put(0,0){\includegraphics[width=\unitlength,page=1]{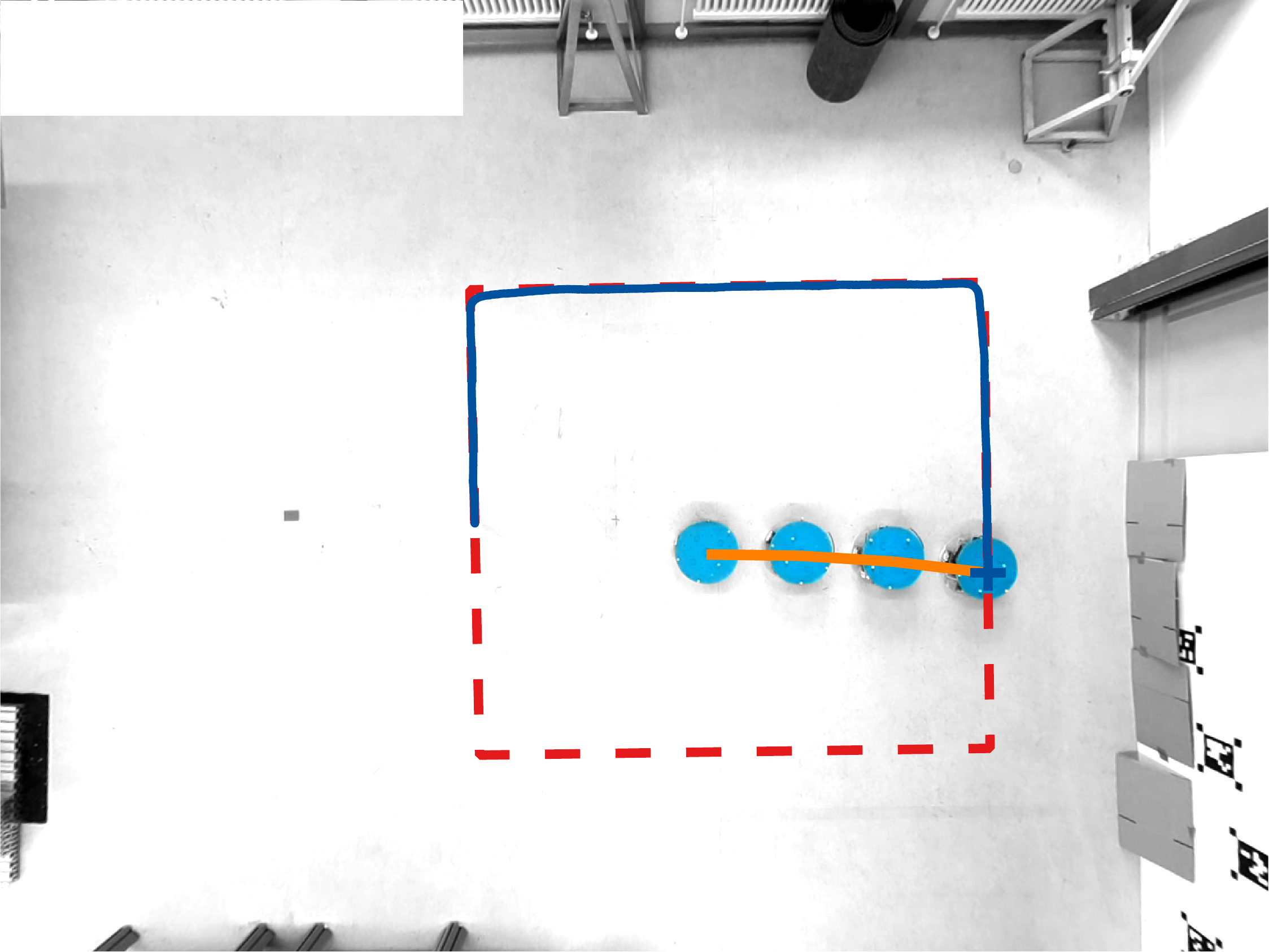}}%
    \put(0.0118159,0.67814321){\makebox(0,0)[lt]{\lineheight{1.25}\smash{\begin{tabular}[t]{l}$t = 45.00\,\textnormal{s}$\end{tabular}}}}%
    \put(0,0){\includegraphics[width=\unitlength,page=2]{exp_ADMM_3.pdf}}%
  \end{picture}%
\endgroup%
\hfill%
	\def\svgwidth{0.4977\columnwidth}%
\begingroup%
  \makeatletter%
  \providecommand\color[2][]{%
    \errmessage{(Inkscape) Color is used for the text in Inkscape, but the package 'color.sty' is not loaded}%
    \renewcommand\color[2][]{}%
  }%
  \providecommand\transparent[1]{%
    \errmessage{(Inkscape) Transparency is used (non-zero) for the text in Inkscape, but the package 'transparent.sty' is not loaded}%
    \renewcommand\transparent[1]{}%
  }%
  \providecommand\rotatebox[2]{#2}%
  \newcommand*\fsize{\dimexpr\f@size pt\relax}%
  \newcommand*\lineheight[1]{\fontsize{\fsize}{#1\fsize}\selectfont}%
  \ifx\svgwidth\undefined%
    \setlength{\unitlength}{1078.25033569bp}%
    \ifx\svgscale\undefined%
      \relax%
    \else%
      \setlength{\unitlength}{\unitlength * \real{\svgscale}}%
    \fi%
  \else%
    \setlength{\unitlength}{\svgwidth}%
  \fi%
  \global\let\svgwidth\undefined%
  \global\let\svgscale\undefined%
  \makeatother%
  \begin{picture}(1,0.75028964)%
    \lineheight{1}%
    \setlength\tabcolsep{0pt}%
    \put(0,0){\includegraphics[width=\unitlength,page=1]{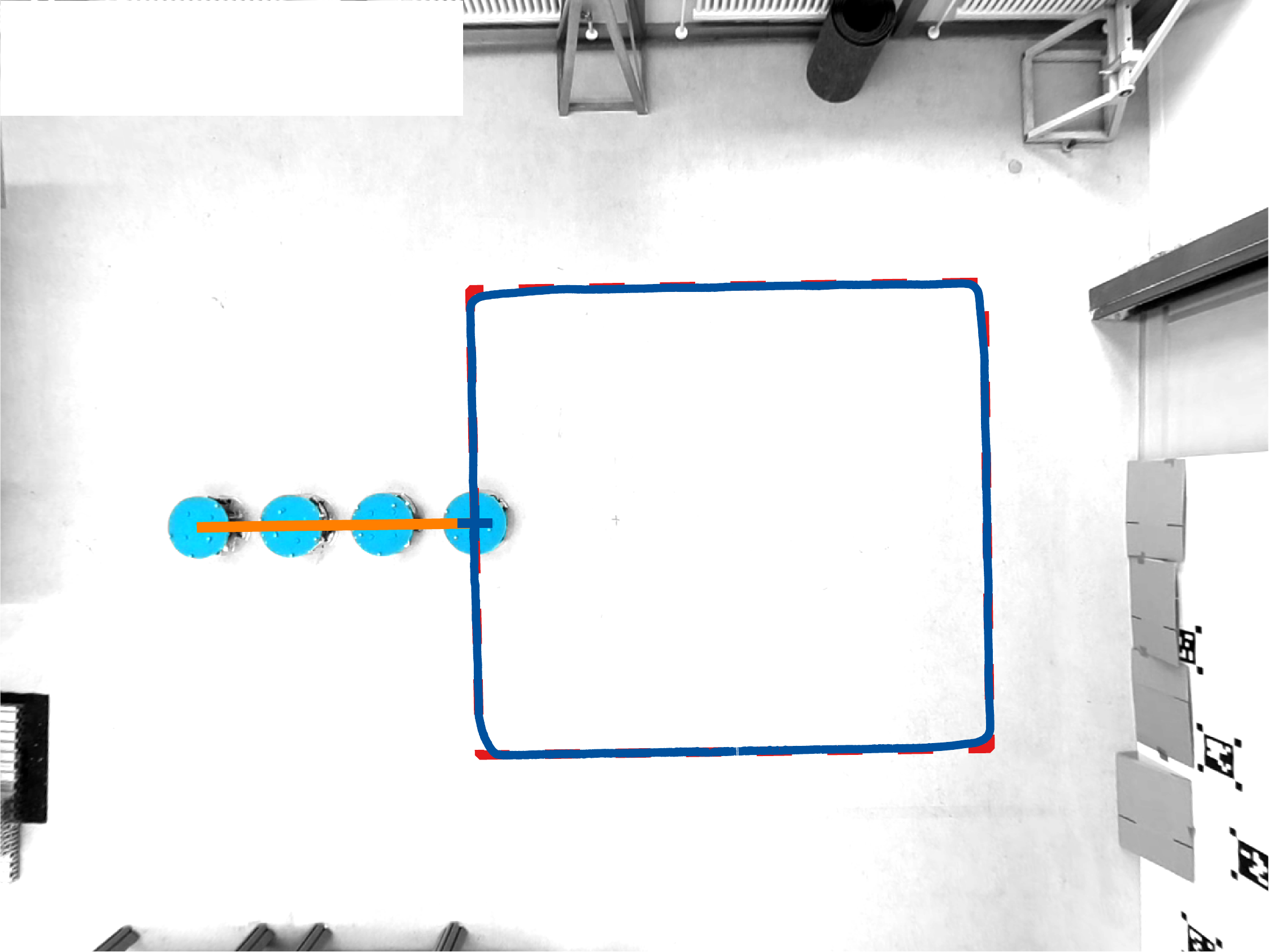}}%
    \put(0.0118159,0.67814321){\makebox(0,0)[lt]{\lineheight{1.25}\smash{\begin{tabular}[t]{l}$t = 75.00\,\textnormal{s}$\end{tabular}}}}%
    \put(0,0){\includegraphics[width=\unitlength,page=2]{exp_ADMM_4.pdf}}%
  \end{picture}%
\endgroup%

	\caption{Rectangle scenario using ADMM-based DMPC: camera images and rectangular path (dashed line), robot one trajectory (solid blue line), and current formation (solid orange line).}
	\label{fig:exp_admm}
\end{figure}

\begin{figure}[t]
	\includegraphics[width=\columnwidth]{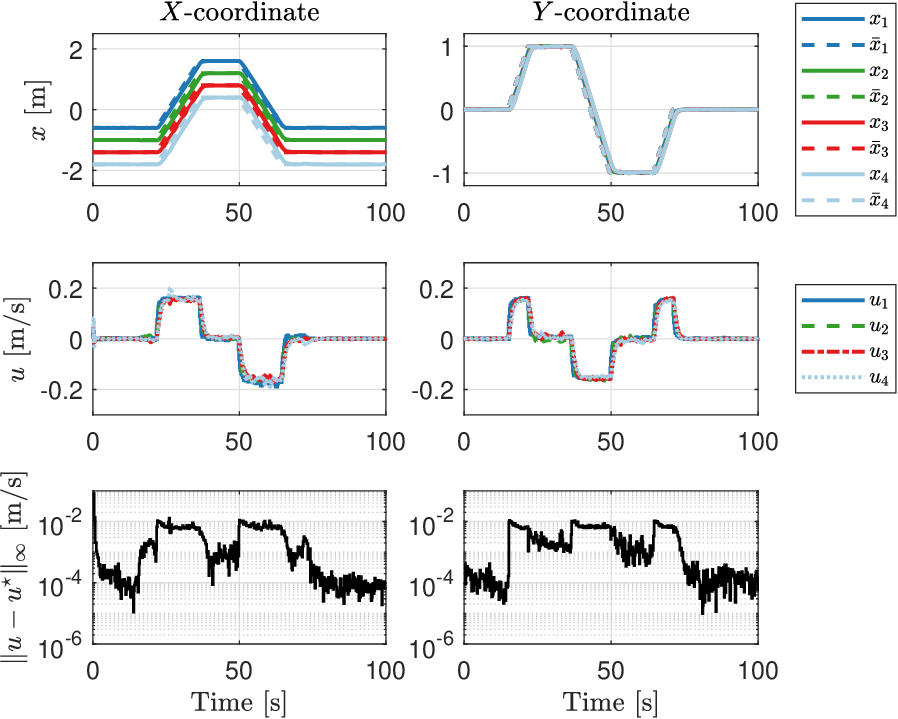}
	\caption{Rectangle scenario using ADMM-based DMPC: closed-loop position $x$ and velocity $u$ trajectories and optimizer residual.}
	\label{fig:admm}
\end{figure}

For ADMM, we empirically tune $\rho$ in the range $\{0.1, 1, 10, 100\}$ for fastest convergence to the OCP minimizer and choose $\rho = 1$.
We fix the number of ADMM iterations executed per MPC step to five. 
That is, we set $l_\mathrm{max} = 5$ in Algorithm~\ref{alg:ADMM}.
Moreover, we warm-start ADMM and set $\bar{z}^0$ and $\gamma^0$ to the solution obtained in the prior MPC step shifted by the time index.
In the first MPC step, we set $\bar{z}^0 = \gamma^0 = 0$.

A video recording of the experiment is available in the digital supplementary material of the article and selected still images from the experiment are shown in Figure~\ref{fig:exp_admm}.
The camera videos and still images are undistorted via suitable image processing algorithms. 
The robot positions are superimposed onto the images using mapped data recorded by the tracking system.
Figure~\ref{fig:exp_admm} shows the robot positions in the $X,Y-$plane, but not the heading angle. This is because the robots are omnidirectional and hence the system dynamics do not depend on the robot orientation.

Figure~\ref{fig:admm} shows the closed-loop state and input trajectories for the individual robots as well as the residual $\|u - u^\star\|_\infty$ between the applied input~$u$ and the centralized optimal solution $u^\star$ obtained from \texttt{IPOPT}~\citep{Wachter2006}.
Plots in the left and right columns of Figure~\ref{fig:admm} present the motion in the $X$- and $Y$-coordinates, respectively.
The plots in the top row show that the control scheme is able to track the desired setpoint accurately.
As can be seen from the bottom plot, ADMM produces input signals which deviate less than $2\cdot 10^{-2}\,\mathrm{m/s}$ from the optimal solution once an initialization phase of a few seconds has passed.
This confirms previous results in the sense that a low number of ADMM iterations per control step can suffice in practice to obtain a reasonable control performance~\citep{vanParys2017,Burk2021,Stomberg2022}.

Table~\ref{tab:admm} shows the median and maximum execution times for ADMM per MPC step and the execution time per ADMM iteration. The data reported therein is based on two experiments with a total duration of approximately three minutes.
The robots took at most $22.27\,\mathrm{ms}$ to execute ADMM per MPC step. Since this is well below the control sampling interval~$\Delta t$, and since we compensate for the computational delay by solving OCP~\eqref{ocp} for $u(t+\Delta t)$, our implementation is sufficiently fast to execute the DMPC scheme in real-time.

\begin{table}\caption{ADMM execution time broken into the steps of Algorithm~\ref{alg:ADMM}. The robots take less than 23\,ms to solve OCP~\eqref{ocp} in each MPC step. This is below the~200\,ms sampling interval and demonstrates the real-time capabilities of the implementation.}\label{tab:admm}
	\centering
	\scalebox{0.95}{
	\begin{tabular}{ l c c}
		& Median\,[ms] & Max.\,[ms]\\
		\hline
		Solve OCP with ADMM & 6.61 & 22.27\\
		Time per ADMM iteration & 0.98 & 16.71\\
		Solve subsystem QP (Step 3) & 0.09 & 14.57 \\
		Communicate $z$ (Steps 4--5) & 0.29 & 11.92\\
		Communicate $\bar{z}$ (Step 6) & 0.25 & 11.63
	\end{tabular}
	}
\end{table}

In addition to the overall ADMM execution time, we are interested in the time required for Steps 3, 4, and 6 of Algorithm~\ref{alg:ADMM}.
Table~\ref{tab:admm} therefore further shows the median and maximum time spans the robots spent on these individual steps.
The time given for Steps 4 and 6 includes the following five substeps: (i) write the variables into an \texttt{LCM} message, (ii) publish the \texttt{LCM} message, (iii) wait for messages sent by neighbors, (iv) write the received variables into local data structures, and (v) if necessary, resend messages to neighbors. 
For Step~4, the time span also includes computing the average in Step~5.
Table~\ref{tab:admm} shows that, in most cases, solving the subsystem QP with \texttt{qpOASES} is significantly faster than exchanging trajectories with neighbors.

As another metric for the execution time, we analyze how much time each robot has spent solving the subsystem QP and communicating/waiting.
Table~\ref{tab:admmRel} reports the time each robot spent in Steps 3 and Steps 4--6 compared to the total time it spent running ADMM. 
All robots spent more time communicating or waiting for their neighbors than solving their quadratic subproblem.
Even though all subproblem QPs have the same structure, robots three and four take less time in Step 3 of Algorithm~\ref{alg:ADMM} than robots one and two.
This is due to performance differences between the workstation computers used to run the DMPC controller for robots three and four compared to robots one and two.

\begin{table}\caption{Fraction of the ADMM execution time spent on Steps 3, 4, and 6 of Algorithm~\ref{alg:ADMM}. Differences between individual robots originate from the varying computational power of the employed hardware.} \label{tab:admmRel}
\centering
\scalebox{0.95}{
\begin{tabular}{ l c c c c}
	Robot & 1 & 2 & 3 & 4\\
	\hline
	Solve subsystem QP (Step 3) & 37\,\% & 33\,\% & 15\,\% & 16\,\% \\
	Communicate (Steps 4--6) & 62\,\% & 66\,\% & 85\,\% & 84\,\%\\
\end{tabular}
}
\end{table}

\subsection{Non-convex Minimum-distance Constraints and dSQP}

The DMPC scheme discussed in the previous subsection relies on cost coupling to meet the formation control specifications.
In this section, we additionally enforce a minimum-distance constraint between neighbors such that for all ${i \in \{2,3,4\}}$ we enforce coupled state constraints
\begin{equation}
	\mathbb{X}_{ij} = \left\{x_i \in \mathbb{R}^{2} \times x_j \in \mathbb{R}^{2} \, \big| \, \left\|x_i - x_{j}\right\|_2^2 \geq d\right\},
    \label{eq:distanceConstraint}
\end{equation} where $j = i-1$ and $d = 0.4\,\mathrm{m}$.
Therefore, OCP~\eqref{ocp}~here is a non-convex quadratically constrained quadratic program~(QCQP), where the distance constraint~\eqref{eq:distanceConstraint} is included in the subsystem inequality constraints~\eqref{nlp:ineq} of robots two to four.

We soften the collision avoidance constraints as described in Remark~\ref{rem:softOCP} and we solve the soft OCP with dSQP.
To this end, we run five SQP iterations per MPC step and three ADMM iterations per SQP iteration.
That is, we set $q_\mathrm{max} = 5$ and ${l_\mathrm{max} = 3}$ in Algorithm~\ref{alg:dSQP}. We again tune $\rho$ in the range $\{0.1,1,10,100\}$ and choose $\rho = 1$.
As for ADMM, we warm-start Algorithm~\ref{alg:dSQP} and set $z^0$ and $\gamma^0$ to the shifted values obtained in the previous MPC step.
We use the Gauss-Newton Hessian approximation $H_i= \nabla_{z_iz_i}^2 f_i$ and therefore omit an initialization for $\nu$ and $\mu$.

\begin{figure}
	\def\svgwidth{0.4977\columnwidth}%
\begingroup%
  \makeatletter%
  \providecommand\color[2][]{%
    \errmessage{(Inkscape) Color is used for the text in Inkscape, but the package 'color.sty' is not loaded}%
    \renewcommand\color[2][]{}%
  }%
  \providecommand\transparent[1]{%
    \errmessage{(Inkscape) Transparency is used (non-zero) for the text in Inkscape, but the package 'transparent.sty' is not loaded}%
    \renewcommand\transparent[1]{}%
  }%
  \providecommand\rotatebox[2]{#2}%
  \newcommand*\fsize{\dimexpr\f@size pt\relax}%
  \newcommand*\lineheight[1]{\fontsize{\fsize}{#1\fsize}\selectfont}%
  \ifx\svgwidth\undefined%
    \setlength{\unitlength}{1078.25033569bp}%
    \ifx\svgscale\undefined%
      \relax%
    \else%
      \setlength{\unitlength}{\unitlength * \real{\svgscale}}%
    \fi%
  \else%
    \setlength{\unitlength}{\svgwidth}%
  \fi%
  \global\let\svgwidth\undefined%
  \global\let\svgscale\undefined%
  \makeatother%
  \begin{picture}(1,0.75028964)%
    \lineheight{1}%
    \setlength\tabcolsep{0pt}%
    \put(0,0){\includegraphics[width=\unitlength,page=1]{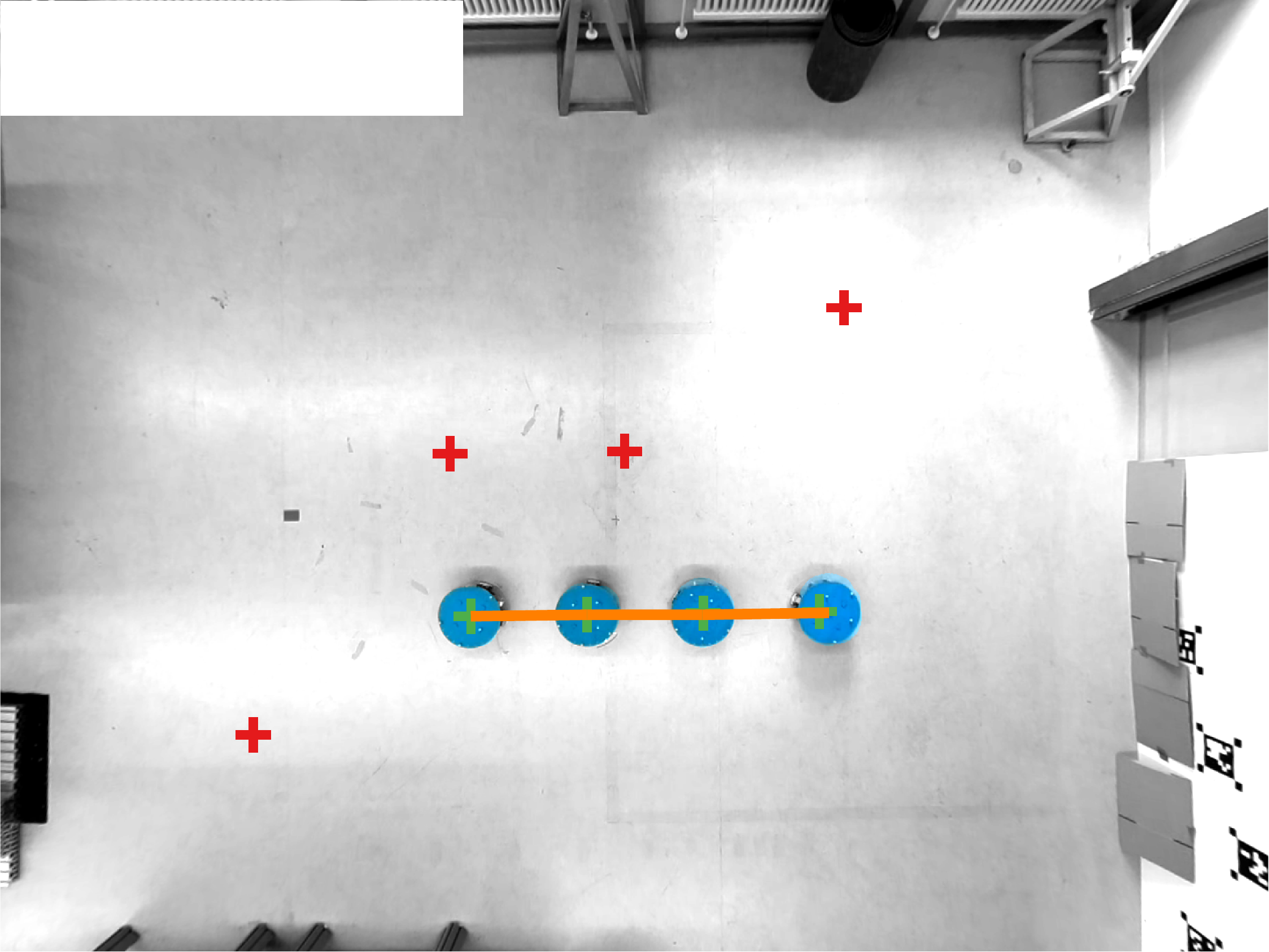}}%
    \put(0.0118159,0.67814321){\makebox(0,0)[lt]{\lineheight{1.25}\smash{\begin{tabular}[t]{l}$t = \phantom{0}0.00\,\textnormal{s}$\end{tabular}}}}%
    \put(0,0){\includegraphics[width=\unitlength,page=2]{exp_dSQP_1.pdf}}%
  \end{picture}%
\endgroup%
\hfill%
	\def\svgwidth{0.4977\columnwidth}%
\begingroup%
  \makeatletter%
  \providecommand\color[2][]{%
    \errmessage{(Inkscape) Color is used for the text in Inkscape, but the package 'color.sty' is not loaded}%
    \renewcommand\color[2][]{}%
  }%
  \providecommand\transparent[1]{%
    \errmessage{(Inkscape) Transparency is used (non-zero) for the text in Inkscape, but the package 'transparent.sty' is not loaded}%
    \renewcommand\transparent[1]{}%
  }%
  \providecommand\rotatebox[2]{#2}%
  \newcommand*\fsize{\dimexpr\f@size pt\relax}%
  \newcommand*\lineheight[1]{\fontsize{\fsize}{#1\fsize}\selectfont}%
  \ifx\svgwidth\undefined%
    \setlength{\unitlength}{1078.25033569bp}%
    \ifx\svgscale\undefined%
      \relax%
    \else%
      \setlength{\unitlength}{\unitlength * \real{\svgscale}}%
    \fi%
  \else%
    \setlength{\unitlength}{\svgwidth}%
  \fi%
  \global\let\svgwidth\undefined%
  \global\let\svgscale\undefined%
  \makeatother%
  \begin{picture}(1,0.75028964)%
    \lineheight{1}%
    \setlength\tabcolsep{0pt}%
    \put(0,0){\includegraphics[width=\unitlength,page=1]{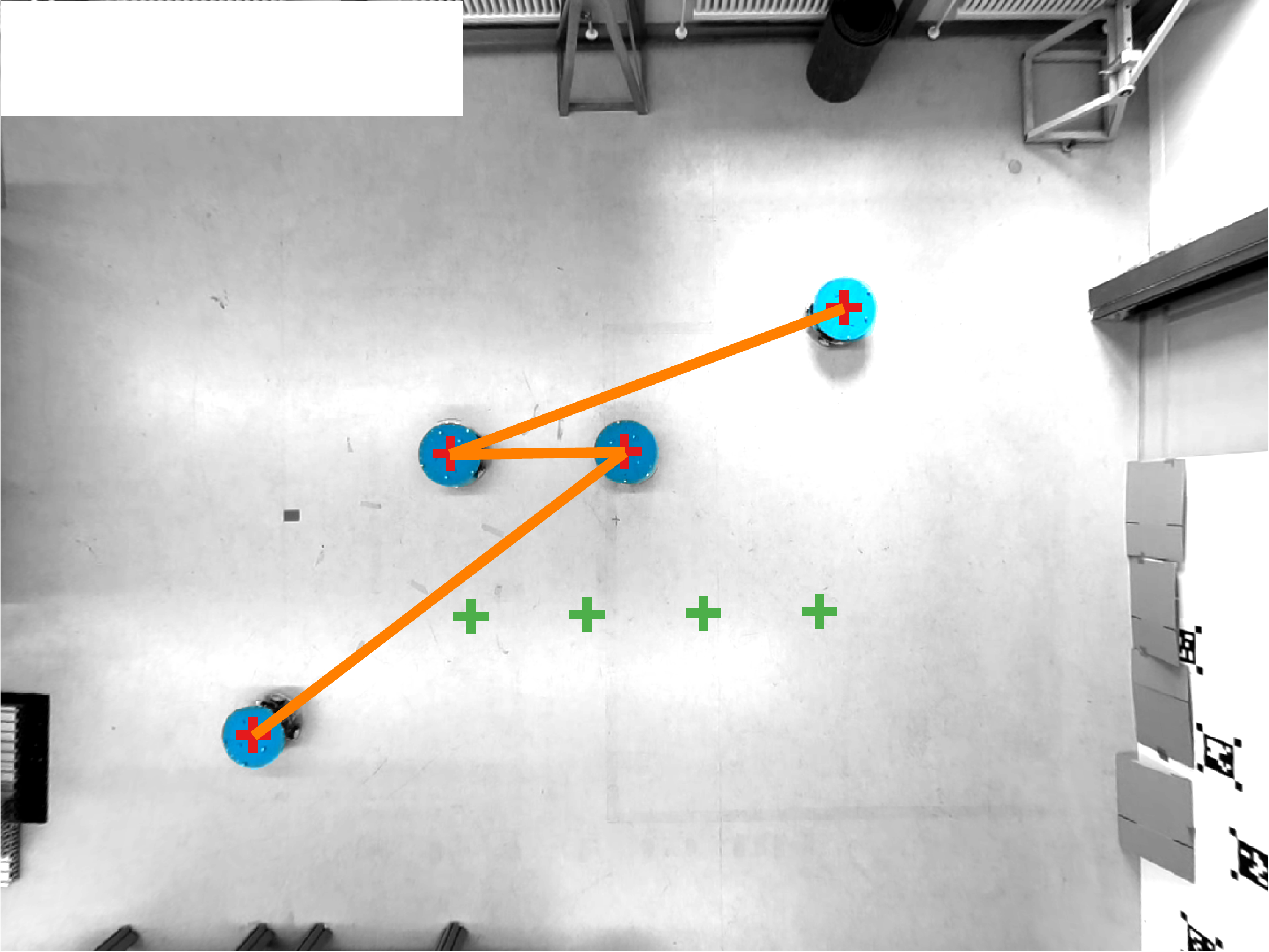}}%
    \put(0.0118159,0.67814321){\makebox(0,0)[lt]{\lineheight{1.25}\smash{\begin{tabular}[t]{l}$t = 10.00\,\textnormal{s}$\end{tabular}}}}%
    \put(0,0){\includegraphics[width=\unitlength,page=2]{exp_dSQP_2.pdf}}%
  \end{picture}%
\endgroup%
\\%
	\def\svgwidth{0.4977\columnwidth}%
\begingroup%
  \makeatletter%
  \providecommand\color[2][]{%
    \errmessage{(Inkscape) Color is used for the text in Inkscape, but the package 'color.sty' is not loaded}%
    \renewcommand\color[2][]{}%
  }%
  \providecommand\transparent[1]{%
    \errmessage{(Inkscape) Transparency is used (non-zero) for the text in Inkscape, but the package 'transparent.sty' is not loaded}%
    \renewcommand\transparent[1]{}%
  }%
  \providecommand\rotatebox[2]{#2}%
  \newcommand*\fsize{\dimexpr\f@size pt\relax}%
  \newcommand*\lineheight[1]{\fontsize{\fsize}{#1\fsize}\selectfont}%
  \ifx\svgwidth\undefined%
    \setlength{\unitlength}{1078.25033569bp}%
    \ifx\svgscale\undefined%
      \relax%
    \else%
      \setlength{\unitlength}{\unitlength * \real{\svgscale}}%
    \fi%
  \else%
    \setlength{\unitlength}{\svgwidth}%
  \fi%
  \global\let\svgwidth\undefined%
  \global\let\svgscale\undefined%
  \makeatother%
  \begin{picture}(1,0.75028964)%
    \lineheight{1}%
    \setlength\tabcolsep{0pt}%
    \put(0,0){\includegraphics[width=\unitlength,page=1]{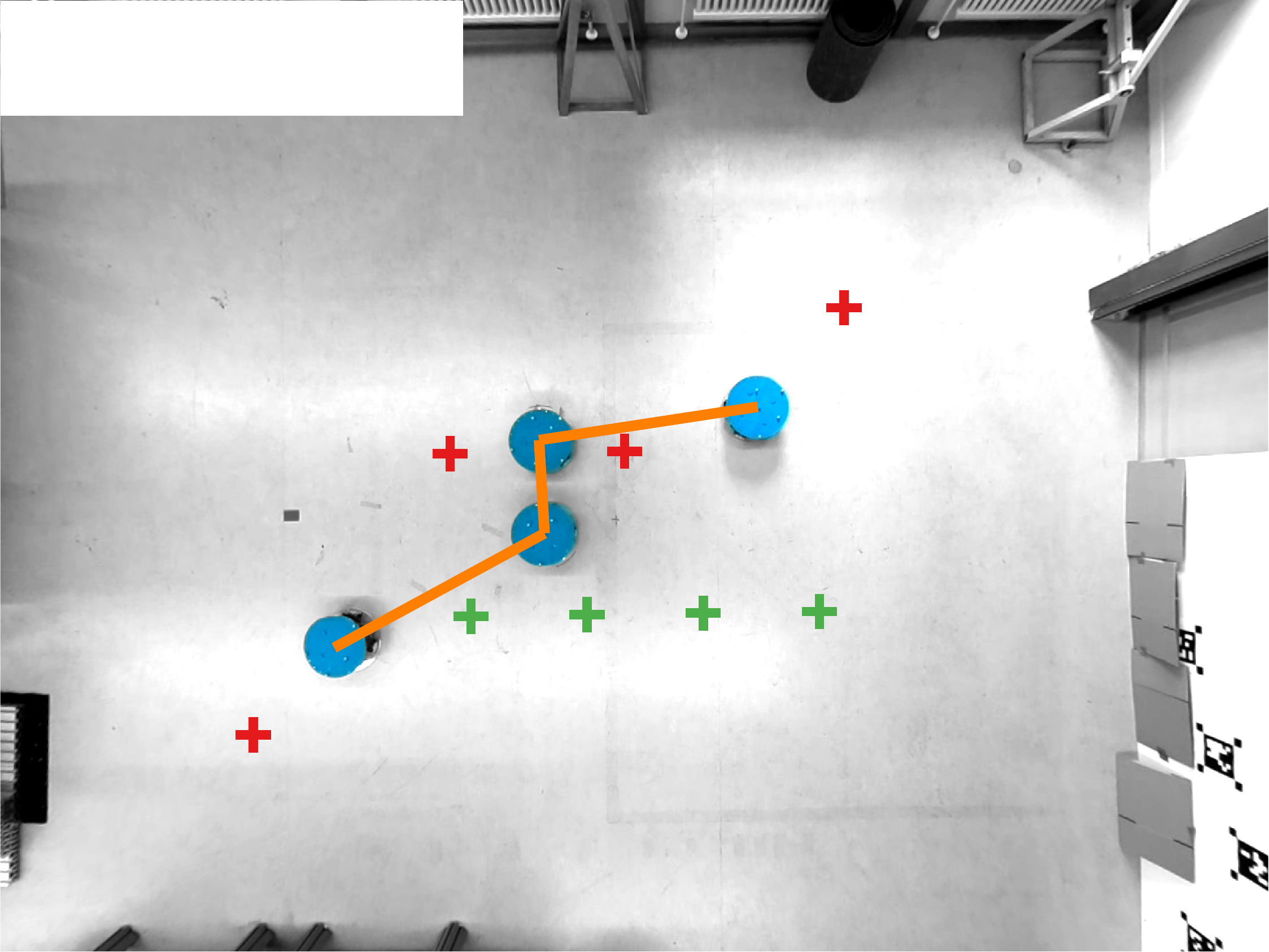}}%
    \put(0.0118159,0.67814321){\makebox(0,0)[lt]{\lineheight{1.25}\smash{\begin{tabular}[t]{l}$t = 22.00\,\textnormal{s}$\end{tabular}}}}%
    \put(0,0){\includegraphics[width=\unitlength,page=2]{exp_dSQP_3.pdf}}%
  \end{picture}%
\endgroup%
\hfill%
	\def\svgwidth{0.4977\columnwidth}%
\begingroup%
  \makeatletter%
  \providecommand\color[2][]{%
    \errmessage{(Inkscape) Color is used for the text in Inkscape, but the package 'color.sty' is not loaded}%
    \renewcommand\color[2][]{}%
  }%
  \providecommand\transparent[1]{%
    \errmessage{(Inkscape) Transparency is used (non-zero) for the text in Inkscape, but the package 'transparent.sty' is not loaded}%
    \renewcommand\transparent[1]{}%
  }%
  \providecommand\rotatebox[2]{#2}%
  \newcommand*\fsize{\dimexpr\f@size pt\relax}%
  \newcommand*\lineheight[1]{\fontsize{\fsize}{#1\fsize}\selectfont}%
  \ifx\svgwidth\undefined%
    \setlength{\unitlength}{1078.25033569bp}%
    \ifx\svgscale\undefined%
      \relax%
    \else%
      \setlength{\unitlength}{\unitlength * \real{\svgscale}}%
    \fi%
  \else%
    \setlength{\unitlength}{\svgwidth}%
  \fi%
  \global\let\svgwidth\undefined%
  \global\let\svgscale\undefined%
  \makeatother%
  \begin{picture}(1,0.75028964)%
    \lineheight{1}%
    \setlength\tabcolsep{0pt}%
    \put(0,0){\includegraphics[width=\unitlength,page=1]{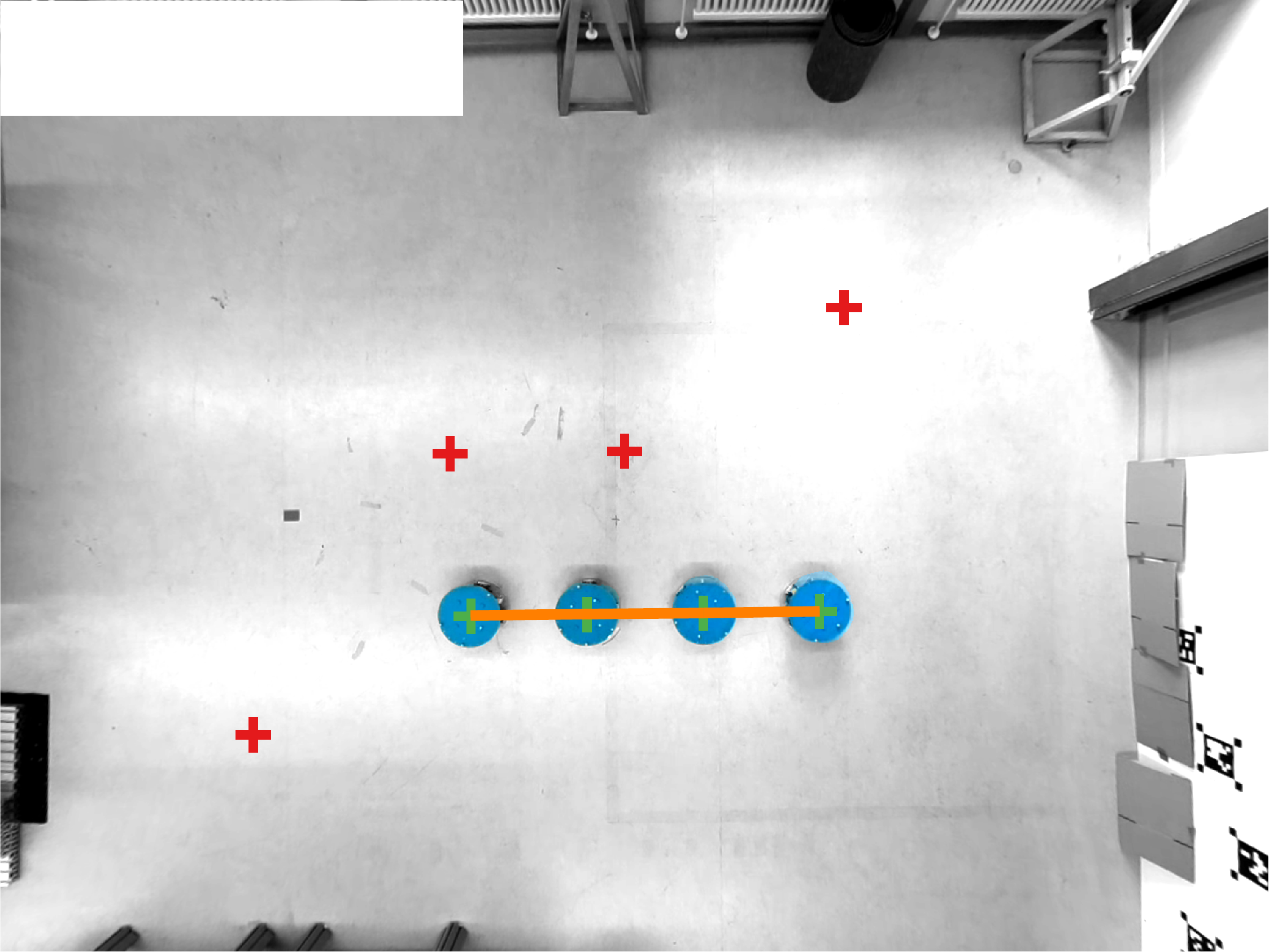}}%
    \put(0.0118159,0.67814321){\makebox(0,0)[lt]{\lineheight{1.25}\smash{\begin{tabular}[t]{l}$t = 30.00\,\textnormal{s}$\end{tabular}}}}%
    \put(0,0){\includegraphics[width=\unitlength,page=2]{exp_dSQP_4.pdf}}%
  \end{picture}%
\endgroup%

	\caption{Formation change scenario using dSQP-based DMPC: camera images and formation one (green crosses), formation two (red crosses), and current formation (solid orange line).}
	\label{fig:exp_dSQP}
\end{figure}

\begin{figure}[t]
	\includegraphics[width=\columnwidth]{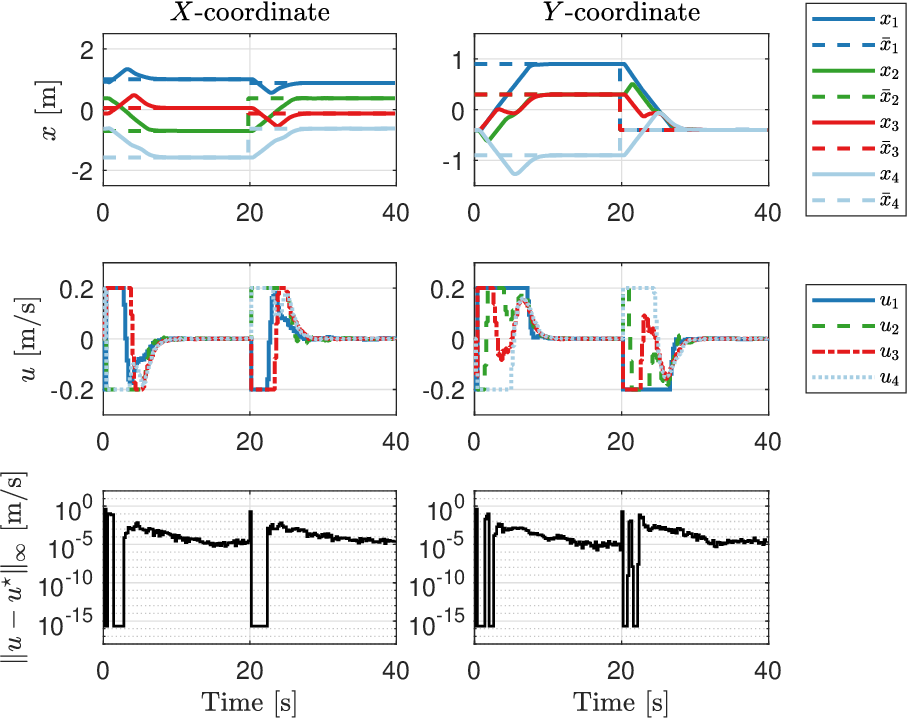}
	\caption{Formation change scenario using dSQP-based DMPC: closed-loop position $x$ and input $u$ trajectories and optimizer residual.}
	\label{fig:dsqp}
\end{figure}

A video recording of the experiment is provided in the digital supplementary material and still images of the experiment are depicted in Figure~\ref{fig:exp_dSQP}. 
Figure~\ref{fig:dsqp} shows closed-loop state and input trajectories as well as the dSQP input residual. 
In the first~20 seconds of the experiment, the robots leave the chain formation and spread across the room.
Then, from $t=20\,\mathrm{s}$ onwards, they return to the chain formation.
This maneuver is more challenging than driving along the rectangular shape, as it requires robots two and three to switch their position while avoiding a collision.
In fact, the shown maneuver cannot be carried out using the ADMM-based DMPC scheme without distance constraints presented in the previous section since simulations indicate that robots two and three would collide.

The bottom plots in Figure~\ref{fig:dsqp} show the residual $\| u - u^\star\|_\infty$ between the applied input $u$ determined by dSQP and the local optimum $u^\star$ found when solving the OCP in centralized fashion via \texttt{IPOPT}.
The control input found by dSQP differs significantly from the \texttt{IPOPT} solution when the experiment begins and when the setpoint changes. 
This is because we lack a good warm-start in these situations. 
However, dSQP returns nearly optimal inputs during the transient phases and upon convergence of the controlled system. 

Table~\ref{tab:dsqp} summarizes the execution times measured in twelve experiments with a total duration of 12 minutes.
The OCP was solved by dSQP in less than $54\,\mathrm{ms}$ and the implementation is hence sufficiently fast.
The table further presents the execution time for the most expensive steps of Algorithm~\ref{alg:dSQP}.
Since the Gauss-Newton Hessian approximation $H_i = \nabla_{z_iz_i}^2f_i$ can be evaluated offline, Step 3 of Algorithm~\ref{alg:dSQP} is cheap. In fact, Algorithm~\ref{alg:dSQP} spends most time executing ADMM in Steps 5--13.

\begin{table}\caption{dSQP execution time broken into the steps of Algorithm~\ref{alg:dSQP}. The robots take less than 54\,ms to solve OCP~\eqref{ocp} in each MPC step. This is well below the 200\,ms sampling interval and demonstrates the real-time capabilities of the implementation.}\label{tab:dsqp}
	\centering
	\scalebox{0.95}{
	\begin{tabular}{ l c c}
		& Median\,[ms] & Max.\,[ms]\\
		\hline
		Solve OCP with dSQP & 33.81 & 53.06\\
		Time per dSQP iteration & 6.23 & 24.57\\
		Build subsystem QP (Step 3) & 0.30 & 1.09\\
		Run ADMM per dSQP iter & 5.99 & 24.29\\
		Time per ADMM iteration & 2.01 & 15.37\\	
		Solve subsystem QP (Step 6) & 0.81 & 12.40 \\
		Communicate $z$ (Step 7--8) & 0.64 & 12.17\\
		Communicate $\bar{z}$ (Step 9) & 0.32 & 11.53
	\end{tabular}
}
\end{table}

Table~\ref{tab:dsqpRel} shows the fraction of the dSQP execution time each robot spent on building and solving the subsystem QP and on communicating with neighbors.
Compared to Table~\ref{tab:admmRel}, robots two to four spent more time in Table~\ref{tab:dsqpRel} on solving the subsystem QP due to the added distance constraint.
In particular, robot two spent more time solving the subsystem QP than communicating or waiting.

\begin{table}\caption{Fraction of the dSQP execution time spent on Steps 3, 6, 7, and 9 of Algorithm~\ref{alg:dSQP}. Due to the minimum-distance constraint and comparatively weaker hardware, robot two spends more time solving the subsystem QPs than communicating trajectories to robots one and three.}\label{tab:dsqpRel}
	\centering
	\scalebox{0.95}{
	\begin{tabular}{ l c c c c}
		Robot & 1 & 2 & 3 & 4\\
		\hline
		Build subsystem QP (Step 3)& 5\,\% & 5\,\% & 3\,\% & 4\,\%\\
		Solve subsystem QP (Step 6)& 34\,\% & 68\,\% & 42\,\% & 29\,\% \\
		Communicate (Steps 7--9) & 60\,\% & 26\,\% & 55\,\% & 65\,\%\\
	\end{tabular}
}
\end{table}

\subsection{Discussion}
The experiments demonstrate the effectiveness of the controller design and of our implementation and the results show that control sampling intervals in the millisecond range can be achieved by cooperative DMPC schemes.
Depending on the OCP structure and the optimization method, both subsystem computations and communication steps can take a substantial part of the execution time.
This is in contrast to~\citep{vanParys2017}, where subsystem computations are a clear bottleneck, as well as to~\citep{Burk2021}, where communication is holding up.
Warm-starting the optimization methods in DMPC allows to find a reasonable control input with only few optimizer iterations per MPC step.

Despite the promising performance in the experiments with respect to control performance and execution time, there are limitations to our approach.
First, we lack rigorous recursive feasibility and convergence guarantees for the control loop because we do not design tailored terminal ingredients for the OCP.
To obtain such theoretial guarantees, one could follow the design procedure presented in~\citep{Darivianakis2019}. 
However, this would only be possible for the scenario without the non-convex distance constraints, as the design in~\citep{Darivianakis2019} only applies to subsystems coupled through polytopic constraints. 
Second, the employed dSQP variant is not guaranteed to converge to an OCP minimizer as we fix the number of ADMM iterations executed per SQP iteration instead of terminating ADMM dynamically.
To obtain local convergence guarantees, one could incorporate a stopping criterion into Algorithm~\ref{alg:dSQP}~\citep{Stomberg2022a}.
This would imply varying optimizer iterations per control step and the additional communication of convergence flags.
Third, our implementation and computing hardware is not hard real-time capable.
That is, even though the observed OCP solution times were strictly shorter than the control sampling intervals, we cannot formally guarantee a sufficiently fast execution of our optimization methods.
Hard real-time guarantees could be obtained by implementing the approaches for execution on microcontrollers or field programmable gate arrays~\citep{Mcinerney2018}.

\section{Conclusion}\label{sec:conclusion}

This article presented experimental results for iterative DMPC schemes based on ADMM and dSQP as well as insights into their efficient implementation for real-time execution.
The decentralized implementation controlled a team of four mobile robots moving in formation.
A dedicated stationary computer was assigned to each robot to run the DMPC application.
The employed decentralized optimization methods ADMM and dSQP solved the centralized OCP sufficiently fast and the DMPC scheme compensated for the computational delay.
In particular, dSQP solved non-convex distance constraints online to avoid collisions between robots.
Future work will consider control schemes with formal stability guarantees, an implementation with hard real-time capabilities, and robots with more sophisticated dynamics, e.g., due to non-holonomic constraints. 

\section*{Acknowledgements}
We thank Carlo Cappello and Petr Listov for their openly available source code, which has been used for reading matrices from disk and for converting between \texttt{CasADi} and \texttt{Eigen} matrices, respectively.

\bibliographystyle{elsarticle-harv} 
\bibliography{paper.bib}

\end{document}